\documentclass[11pt]{article}
\usepackage{jcapmod}

\usepackage{booktabs}
\usepackage[english]{babel}
\usepackage{amsmath,amssymb,amsbsy,amstext, amsthm}
\usepackage{hyperref}
\usepackage{graphicx}
\usepackage{amsfonts}
\usepackage{amssymb}
\usepackage[small]{caption}	
\usepackage{upgreek}
\usepackage{framed}
\usepackage{wrapfig}
\usepackage{multirow}
\usepackage{subfig}
\usepackage{bbm}
\usepackage{bm}

\definecolor{brown(traditional)}{rgb}{0.59, 0.29, 0.0}

\usepackage[usenames]{xcolor}

\usepackage{colortbl}
\definecolor{lightgreen}{cmyk}{0.2, 0, 0.2, 0.2}
\definecolor{lightgray}{cmyk}{0.1,0.2,0,0.1}
\definecolor{lightgray2}{cmyk}{0.1,0.1,0,0.1}

\definecolor{NewRed}{RGB}{200,37,6}
\definecolor{NewOrange}{RGB}{222,106,16}
\definecolor{NewGreen}{RGB}{0,136,43}

\makeatletter
\newlength{\apb@width}
\newcommand{\autoparbox}[2][c]{\settowidth{\apb@width}{#2}\parbox[#1]{\apb@width}{#2}}

\makeatother

\makeatletter
\newcommand{\raisemath}[1]{\mathpalette{\raisem@th{#1}}}
\newcommand{\raisem@th}[3]{\raisebox{#1}{$#2#3$}}
\makeatother

\numberwithin{equation}{section}

\def\beq{\begin{equation}}
\def\eeq{\end{equation}}

\def\bea{\begin{eqnarray}}
\def\eea{\end{eqnarray}}

\def\Beq{\begin{equation}\begin{aligned}}
\def\Eeq{\end{aligned}\end{equation}}

\def\d{{\rm d}}
\def\dd{{\rm d}}

\def\beq{\begin{equation}}
\def\eeq{\end{equation}}
\def\bea{\begin{eqnarray}}
\def\eea{\end{eqnarray}}

\def\I{{\sf I}}
\def\J{{\sf J}}
\def\K{{\sf K}}
\def\L{{\sf L}}

\def\d{{\rm d}}
\def\dd{{\rm d}}

\def\d{{\bar R}}

\def\eps{\boldsymbol\varepsilon}

\def\Mp{M_{\rm pl}}
\def\d{{\rm d}}

\def\k{{\bf k}}

\def\q{{\bf q}}

\def\x{{\bf x}}
\def\y{{\bf y}}
\def\0{{\bf 0}}

\def\z{{\bf z}}

\def\hs{\hskip 1pt}

\newcommand{\nn}{\nonumber \\ }

\newcommand{\re}{ {\rm Re} }

\DeclareRobustCommand{\SkipTocEntry}[4]{}

\begin{document}

\begin{titlepage}

\setcounter{page}{1} \baselineskip=15.5pt \thispagestyle{empty}

\bigskip\

\vspace{1cm}
\begin{center}
{\fontsize{20}{24}\selectfont  \sffamily \bfseries  Partially Massless Fields During Inflation}
\end{center}

\vspace{0.2cm}
\begin{center}
{\fontsize{13}{30}\selectfont  Daniel Baumann,$^{1}$ Garrett Goon,$^{1,2}$ Hayden Lee,$^{3,4}$  and Guilherme L. Pimentel$^{1}$} 
\end{center}

\begin{center}

\vskip 8pt
\textsl{$^1$  Institute of Theoretical Physics, University of Amsterdam,\\Science Park 904, Amsterdam, 1098 XH, The Netherlands}

\vskip 8pt
\textsl{$^2$ Institute for Theoretical Physics, 
Utrecht University, \\ Princetonplein 5, Utrecht, 3584 CC, The Netherlands}

\vskip 8pt
\textsl{$^{3}$ Department of Physics, Harvard University,\\ Cambridge, MA 02138, USA}  

\vskip 8pt
\textsl{$^{4}$ Institute for Advanced Study, \\ Hong Kong University of Science and Technology, Hong Kong}  
\vskip 7pt

\end{center}

\vspace{1.2cm}
\hrule \vspace{0.3cm}
\noindent {\sffamily \bfseries Abstract} \\[0.1cm]
The representation theory of de Sitter space allows for a category of partially massless particles which have no flat space analog, but could have existed during inflation.  We study the couplings of these exotic particles to inflationary perturbations and determine the resulting signatures in cosmological correlators. When inflationary perturbations interact through the exchange of these fields, their correlation functions inherit scalings that cannot be mimicked by extra massive fields. We discuss in detail the squeezed limit of the tensor-scalar-scalar bispectrum, and show that certain partially massless fields can violate the tensor consistency relation of single-field inflation. We also consider the collapsed limit of the scalar trispectrum, and find that the exchange of partially massless fields enhances its magnitude, while giving no contribution to the scalar bispectrum. These characteristic signatures provide clean detection channels for partially massless fields during inflation. 
 
\vskip 10pt
\hrule
\vskip 10pt

\vspace{0.6cm}
 \end{titlepage}

\tableofcontents

\newpage
\section{Introduction}

 Observations suggest that our Universe started as de Sitter space~\cite{Ade:2015lrj} and will end as de Sitter space~\cite{Riess:1998cb, Perlmutter:1998np}. Understanding the physics of de Sitter (dS) is therefore of particular relevance~\cite{Spradlin:2001pw,Anninos:2012qw,Akhmedov:2013vka}. There exist a number of interesting features of this spacetime which do not have counterparts in flat space.  For example, while particles in Minkowski are either massive or massless~\cite{Bargmann:1948ck,Wigner:1939cj}, the representation theory of de Sitter space allows for an extra category of partially massless (PM) particles~\cite{Deser:2003gw,Newton,Thomas}. At special discrete values of the mass-to-Hubble ratio, $m/H$, the theory gains an additional gauge symmetry and some of the lowest helicity modes of the would-be massive particles become pure gauge modes. In this paper, we revisit the theoretical and observational status of partially massless particles during inflation. 

\vskip 4pt
 Partially massless particles have a number of intriguing features that motivate us to study their effects during inflation, despite their somewhat exotic nature and their uncertain status as interacting quantum field theories.\footnote{In order to write a local theory with the right number of degrees of freedom, one imposes a gauge symmetry that must be obeyed by the higher-spin Lagrangian. It turns out to be hard to find self-consistent interacting theories of these degrees of freedom. It might be that a single light degree of freedom in the spectrum implies the existence of many other higher-spin degrees of freedom.  An extreme example of this is the occurrence of infinitely many fields in the Vasiliev theory of massless higher-spin particles in (A)dS~\cite{Vasiliev:1990en, Vasiliev:1992av}. Similarly, interacting theories of an infinite tower of partially massless fields were studied~in~\cite{Bekaert:2013zya, Alkalaev:2013fsa, Alkalaev:2014nsa, Joung:2015jza, Brust:2016zns}.} 
First of all, if PM particles existed during inflation, they would lead to rather distinct imprints in cosmological correlation functions.  Moreover, the chances of detecting these signals may even be bigger than for massive particles, since the rate at which PM particles would be produced during inflation is larger than that for massive particles. Furthermore, while massive fields decay on superhorizon scales, the amplitude of certain PM fields can remain constant or even grow. 
They can therefore survive until the end of inflation and their contributions to the soft limits of inflationary correlators are unsuppressed. Another interesting feature of PM particles is that their masses are protected against radiative corrections by the gauge symmetry. 
Finally, since the existence of PM particles is tied to the non-zero (and nearly constant) Hubble parameter during inflation, their detection would provide further evidence for an inflationary, de Sitter-like period of expansion in the early universe.

\vskip 4pt
To describe the effects of PM particles on cosmological correlators, we will construct gauge-invariant couplings between higher-spin particles and the inflationary scalar and tensor fluctuations. 
 Using these couplings, we will show that partially massless higher-spin particles leave unique imprints in the soft limits of inflationary correlation functions.
 These soft limits are a well-known detection channel for extra fields during inflation since a symmetry fixes their form in single-field inflation~\cite{Maldacena:2002vr,Creminelli:2004yq}.
Most of the focus, so far, has been on the soft limit of the scalar bispectrum  $\langle \zeta \zeta \zeta \rangle$, where $\zeta$ is the primordial curvature perturbation, which would receive characteristic {\it non-analytic} contributions from massive particles~\cite{Chen:2009we,Chen:2009zp,Noumi:2012vr,Arkani-Hamed:2015bza, Mirbabayi:2015hva, Chen:2015lza, Lee:2016vti}. A {\it strict} violation of the consistency relation---that is, a modification of the leading term in the squeezed limit---would require extra massless scalars.  

\vskip 4pt
An even more robust consistency relation exists for the soft limit of $\langle \gamma \zeta \zeta \rangle$, where $\gamma$ is a tensor fluctuation. 
In that case, the leading term in the tensor squeezed limit cannot be altered by the addition of light scalars. Moreover, contributions from massive spin-$s$ fields are constrained by the Higuchi-Deser-Waldron (HDW) bound in dS space, $m^2\ge s(s-1)H^2$~\cite{Higuchi:1986py, Deser:2001xr, Zinoviev:2001dt}. Achieving a strict violation of the tensor consistency relation has so far been restricted to models where a subset of the de Sitter isometries are fully broken. This includes models with large corrections to the kinetic terms of the spinning fields\footnote{It is possible to evade the HDW bound by writing quadratic actions that are non-covariant, but respect the preferred slicing of the spacetime during inflation. If the non-covariant  terms are comparable in magnitude to the covariant terms, then the signs of the kinetic terms for the different degrees of freedom of the higher-spin field can be tuned separately.  
In this case, one can protect lower-helicity modes from becoming ghost-like, while allowing for mass terms that evade the unitarity bound. In other words, by assuming a large breaking of de Sitter symmetry, there is enough freedom to write ghost-free quadratic actions that violate the HDW bound. We thank Andrei Khmelnitsky for discussions on this point~\cite{Higviolation}.}~\cite{Kehagias:2017cym,Higviolation}  and models with broken spatial isometries due to position-dependent background fields~\cite{Endlich:2012pz, Piazza:2017bsd}.

\vskip 4pt
In this paper, we will show that the presence of PM fields during inflation can lead to a strict violation of the consistency relation for $\langle \gamma \zeta \zeta \rangle$ with a characteristic angular dependence. In addition, the exchange of PM particles creates an enhanced scalar trispectrum $\langle \zeta \zeta \zeta \zeta \rangle$, without producing a scalar bispectrum $\langle \zeta \zeta \zeta \rangle$. These features of the cosmological correlators are rather unique and provide a clean detection channel for PM fields during inflation.

\vskip 4 pt
\paragraph{Outline} The paper is organized as follows. In Section \ref{sec:HS}, we review the higher-spin representations in de Sitter space, and derive the interaction vertices with the inflationary scalar and tensor fluctuations. In Section \ref{sec:CC}, we determine the imprints of partially massless fields in inflationary correlators, with particular emphasis on the tensor-scalar-scalar bispectrum $\langle \gamma \zeta \zeta\rangle$ and the scalar trispectrum $\langle \zeta \zeta \zeta \zeta \rangle$. In Section \ref{sec:Conc}, we present our conclusions. 
The appendices contain technical details of the computations presented in the main text. In Appendix~\ref{app:A}, we expand on the free theory of higher-spin fields in de Sitter space. In Appendix~\ref{app:coupling}, we derive the interaction vertices between a spin-4 field and the inflationary fluctuations, and discuss generalizations to higher spin. In Appendix~\ref{app:corr}, we present the calculation of $\langle \gamma \zeta \zeta\rangle$ and $\langle \zeta \zeta \zeta \zeta \rangle$ in the in-in formalism. Finally, in Appendix \ref{app:OPE}, we derive the scaling behavior of the collapsed trispectrum and the squeezed bispectrum using the operator product expansion and the wavefunction of the universe. 

 \vskip 4pt
\paragraph{Notation and conventions} 
Throughout the paper, we use natural units, $c=\hbar=1$, with reduced Planck mass $\Mp^2=1/8\pi G$. The metric signature is (${-}\hs{+}\hs{+}\hs{+}$), spacetime indices are denoted by Greek letters, $\mu, \nu, \cdots =0,1,2,3$, and spatial indices by Latin letters, $i,j,\cdots=1,2,3$. Conformal time is $\eta$, and a prime (overdot) on a field refers to a derivative with respect to conformal (physical) time. Three-dimensional vectors are written in boldface, $\k$, and unit vectors are hatted, $\hat \k$. Expectation values with a prime, $\langle f_{\k_1} f_{\k_2}\rangle' $, indicate that the overall momentum-conserving delta function has been dropped. The spin and depth of a field are labelled by $s$ and~$t$, respectively. Partial and covariant derivatives are written as $\partial_\mu$ and $\nabla_\mu$, respectively, with $\nabla_{\mu_1\cdots\mu_s}\equiv\nabla_{\mu_1}\cdots\nabla_{\mu_s}$.

\newpage
\section{Higher-Spin Fields during Inflation} 
\label{sec:HS}

In this section, we introduce the free theory of higher-spin fields during inflation and their couplings to the inflationary perturbations. 
We begin, in \S\ref{sec:group}, by reviewing the classification of particles in de Sitter space.  We highlight the existence of a class of partially massless fields. The free theory of these fields will be presented in \S\ref{sec:freeth}.  Finally, in \S\ref{sec:couplings}, we discuss the allowed couplings between these fields and the inflationary scalar and tensor fluctuations.

\subsection{De Sitter Representations}
\label{sec:group}

The relativistic equations of motion for particles of arbitrary spin in flat space were derived by Fierz and Pauli in~\cite{Fierz:1939ix}, based on the requirement of positive energy. This is equivalent to the condition that one-particle states transform under unitary irreducible representations of the Poincar\'e group~\cite{Wigner:1939cj, Bargmann:1948ck}. These representations are characterized by the eigenvalues of the two Casimirs of the group: 
\begin{align}
{\cal C}_1 &\equiv P_\mu P^\mu = m^2\, , \\
 {\cal C}_2 &\equiv  W_\mu W^\mu = -s(s+1)\hskip 1pt m^2\, , 
\end{align}
where $P_\mu$ is the four-momentum and $W_\mu$ is the Pauli-Lubanski pseudovector. The mass $m$ of a particle is a non-negative real number, whereas its spin $s$ is a non-negative half integer. We distinguish between massive and massless particles. In four spacetime dimensions, massive particles carry $2s+1$ degrees of freedom (transverse and longitudinal polarizations). For the massless case, the theory gains a gauge symmetry which eliminates the longitudinal degrees of freedom, and only the two transverse polarizations remain.

\vskip 4pt
Similarly, particles in de Sitter space are classified as unitary irreducible representations of the isometry group SO(1,4). 
The Casimirs of the de Sitter group have eigenvalues \cite{Deser:2003gw, deWit:2002vz}
\begin{align}
	{\cal C}_{1} &\equiv \tfrac{1}{2}M_{AB}M^{AB} = m^2-2(s-1)(s+1)H^2 \, , \\[2pt]
	{\cal C}_{2} &\equiv W_A W^A = -s(s+1)\big(m^2-(s^2+s-\tfrac{1}{2})H^2\big) \, ,
\end{align}
where $M_{AB}$ are the generators of SO(1,4) with $A,B \in \{0,\cdots, 4\}$ and $W_A$ is the five-dimensional Pauli-Lubanski pseudovector, constructed out of two Lorentz generators. 
There is no globally timelike Killing vector in de Sitter space. This is clear when describing de Sitter space in embedding coordinates, where all isometry generators correspond to rotations or Lorentz boosts. This implies that the positivity constraints are imposed on certain combinations of mass and spin of the representations, rather than mere positivity of the mass. 
The (non-scalar, bosonic) representations of the de Sitter group fall into three distinct categories~\cite{Thomas, Newton}: 
\begin{center}
\begin{tabular}{ccc}
{\color{NewRed}principal series} & {\color{NewOrange}complementary series} & {\color{NewGreen}discrete series} \\[4pt]
\ \ $\displaystyle \frac{m^2}{H^2} \ge \left(s-\frac{1}{2}\right)^2$ \quad & \quad $\displaystyle s(s-1) < \frac{m^2}{H^2} < \left(s-\frac{1}{2}\right)^2$ \quad & \quad $\displaystyle \frac{m^2}{H^2} = s(s-1)-t(t+1)$\, , 
\end{tabular}
\end{center}
\vskip 4pt
where the label $t=0,1,2, \ldots,s-1$ is called the ``depth'' of the field. Masses that do not belong in one of the above categories correspond to non-unitary representations and are not allowed in the spectrum (see Fig.~\ref{fig:spectrum}). For spinning fields, the complementary series consists of a narrow range of mass values. For the representations in the discrete series, the system has gauge symmetries which remove would-be ghost degrees of freedom from the spectrum.  The discrete series can be further divided into the following two subcategories:

\begin{figure}[t!]
    \centering
      \includegraphics[width=.9\textwidth]{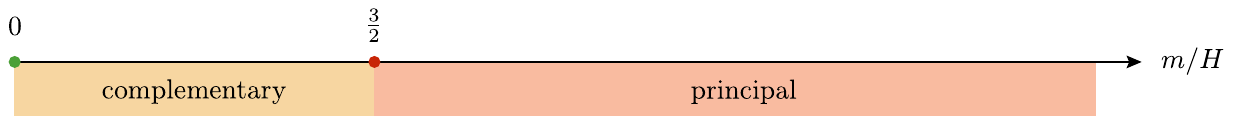}\\[10pt]
         \includegraphics[width=.9\textwidth]{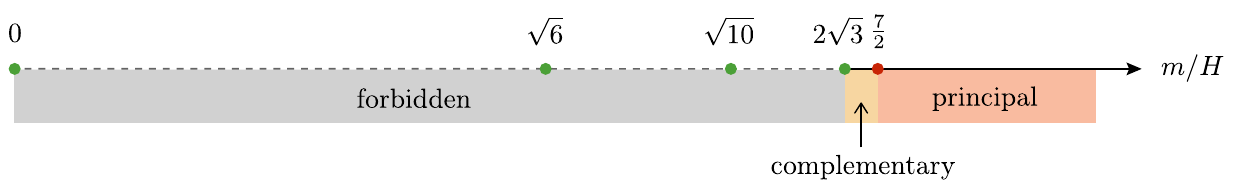}
         \\
    \caption{Spectrum of spin-0 (top) and spin-4 fields (bottom) in de Sitter space. The green points correspond to masses in the discrete series.}
    \label{fig:spectrum}
\end{figure}

\begin{itemize}
\item {\bf Partially massless fields} \hskip 6pt Representations of the discrete series with $t \ne s-1$ correspond to partially massless fields~\cite{Deser:2001pe, Deser:2001us}. These fields share some features of massive and massless fields in flat space. On the one hand, they carry more than two degrees of freedom, akin to massive fields. On the other hand, their correlation functions have power-law behavior, which is similar to that of massless fields.  Unlike massive fields, PM fields may survive until the end of inflation and therefore be directly observable.
To derive this effect, however, one needs to extend the definition of PM fields beyond the de Sitter limit to general FRW cosmologies.\footnote{We emphasize that the irreducible representations of the dS group carried by PM fields have no counterpart in Minkowski space. This can also be understood from the perspective of field theory; only when $m$ takes values that are particular multiples of $H\ne 0$, the action develops new gauge invariance that removes certain lower helicity components. As a result, there is a discontinuity in the ($m,H$)-plane at $(0,0)$ for $s\ge 2$, since the number of degrees of freedom depends on the way we take the limit $H\to 0$.} This remains an open problem (but see App.~\ref{app:A} and Refs.~\cite{Bernard:2017tcg, Aros:2017ror, Cortese:2017ieu, Bekaert:2017bpy}). In this paper, we instead consider the imprints of PM fields through their conversion to massless scalar and tensor perturbations during inflation. This doesn't require us to follow their evolution after inflation.

\item {\bf Massless fields} \hskip 6pt Representations of the discrete series with maximum depth, $t=s-1$, correspond to massless fields.
Interacting theories of massless higher-spin particles are notoriously difficult to construct in flat space, restricted by various powerful no-go theorems~\cite{Weinberg:1964ew, Coleman:1967ad,Weinberg:1980kq} (see~\cite{Bekaert:2010hw, Rahman:2015pzl} for reviews).
In order for particles to remain massless, their interactions need to be protected by a gauge symmetry. Typically, this condition is sufficient to uniquely determine the structure of the nonlinear theory.\footnote{For massless spin-1 particles, the couplings must satisfy a sum rule which corresponds to charge conservation, as in electromagnetism. For a massless spin-2 particle, the couplings to other matter fields must be universal, thus implying the equivalence principle, as in general relativity.}  
The constraints on the couplings of massless higher-spin fields to charged matter imply that these fields cannot induce any long-range forces.  The argument relies heavily on the flat-space S-matrix, and one might wonder whether the same conclusion holds in (A)dS.  It does not: a non-zero cosmological constant allows for the existence of interacting theories of massless higher-spin fields (for reviews, see~\cite{Vasiliev:1999ba, Giombi:2016ejx}).  Having said that, all known examples require an infinite tower of massless higher-spin fields, and their interactions are highly constrained. It is not known if these examples exhaust the list of possibilities for theories of massless higher-spin fields in curved spacetimes. 
\end{itemize}

\noindent
The inflationary phenomenology of massive particles in the principal and complementary series was studied~in~\cite{Arkani-Hamed:2015bza,Baumann:2011nk,Chen:2009zp,Kehagias:2017cym,Lee:2016vti,Noumi:2012vr, Chen:2016hrz,Kumar:2017ecc}, while the observational prospects were considered~in~\cite{Sefusatti:2012ye, Meerburg:2016zdz,Gleyzes:2016tdh, MoradinezhadDizgah:2017szk, MoradinezhadDizgah:2018ssw}. In this paper, we will study novel effects arising from the presence of particles in the discrete series. In particular, we will study the cubic couplings of partially massless fields to the inflationary perturbations and the resulting signatures in cosmological correlators. These give the leading contributions to the cosmological correlators that we consider in this paper, while self-interactions of PM fields do not contribute at tree level.  
The cubic couplings can be made gauge invariant at the desired order, and we will show an explicit construction of this coupling for PM spin-4 particles.

\subsection{Free Theory of Partially Massless Fields}
\label{sec:freeth}

Let us briefly review the free theory of PM fields. Further details are presented in Appendix~\ref{app:A}. The on-shell equations of motion for a spin-$s$, depth-$t$ field are a generalization of the Fronsdal equations for massless particles in de Sitter space~\cite{Fronsdal:1978vb}:
\begin{align}
	\Big[\Box-\big(s+2-t(t+1)\big)H^2\Big]\sigma_{\mu_1\cdots\mu_s}=0\, , \quad 	\nabla^{\mu}\sigma_{\mu\mu_2\cdots \mu_s}=0 \, , \quad {\sigma^{\mu}}_{\mu\mu_2\cdots \mu_s}=0\, .\label{PMeommain}
\end{align}
These equations have a gauge symmetry that reduces the number of degrees of freedom from the naive estimate. For massless fields, this gauge invariance allows us to set all $\sigma_{0 \mu_2 \cdots \mu_s}$ components to zero. In the case of PM fields, only a subset of the timelike components of $\sigma_{\mu_1\cdots \mu_s}$ can be set to zero. The remaining components are labeled by a ``spatial spin" $n$. Moreover, each component has a helicity label, denoted by $\lambda$. We can therefore write the components of a PM field of spin $s$ and with $n$ spatial indices as 
\beq
	\sigma_{i_1\cdots i_n \eta\cdots \eta} = \sum_\lambda\sigma_{n,s}^\lambda\hskip 1pt \varepsilon^\lambda_{i_1\cdots i_n}\, ,
	\label{helicitydecomp}
\eeq
with $s \ge n\ge |\lambda|$ and $|\lambda|=t+1,\cdots,s$.
The conventions for the polarization tensors $\varepsilon^\lambda_{i_1\cdots i_n}$, as well as explicit expressions for the mode functions $\sigma^\lambda_{n,s}$, can be found in Appendix \ref{app:A}.

\vskip 4pt
A few features of the power spectra of fields in the discrete series deserve to be highlighted. 
First of all, the mode function $\sigma^\lambda_{n, s}$ takes the following schematic form: 
\beq
\sigma^\lambda_{n,s} =\frac{1}{\sqrt{2k}}\frac{1}{(-H\eta)^{n-1}}\left[a_0+\cdots+\frac{a_{t}}{(-k\eta)^{t}}\right]e^{-ik\eta}\, ,\label{eq:mode}
\eeq
with constant coefficients $a_n$. Notice that this function is elementary rather than transcendental (as in the generic massive case), being of the form $e^{-i k \eta}$ multiplied by a rational function in $k\eta$. The fact that these mode functions are very similar to flat space mode functions begs for an explanation. In Appendix~\ref{app:A},  we provide a heuristic explanation for this feature of PM fields~\cite{Metsaev:2007fq,Maldacena:2011mk,Metsaev:2014iwa}. The argument shows,  in a sense, why PM fields exist and also suggests a way to extend the definition of PM fields away from the perfect de Sitter limit.  

\vskip 4pt
Second, the two-point function of fields in the discrete series scales at late times as 
\begin{align}
	\langle\sigma_{\mu_1\cdots \mu_s}\sigma^{\mu_1\cdots \mu_s}\rangle 
\xrightarrow{\ \eta\to 0\ } \eta^{2\Delta}\, ,\label{sigma2pt}
\end{align}
where $\Delta = 1-t$ is the conformal dimension of a field with spin $s$ and depth $t$. 
We see that the two-point function freezes at late times for $t=1$ and diverges for $t>1$. It is unclear whether this carries any physical significance, since this two-point function is, of course, gauge dependent. On the one hand, the gauge-invariant curvature tensors built from these higher-spin fields have additional derivatives and their two-point functions therefore vanish at late times. On the other hand, when PM fields are minimally coupled to matter, these late-time divergences can become physical.\footnote{The late-time growth of the power spectrum of certain higher-spin fields is worrisome, as it could be the sign of an instability. In order to determine whether the backreaction on the dS geometry is large, it would be most natural to compute the contribution of the higher-spin fields to the background energy density. However, because the form of the higher-spin action on a generic curved background is unknown, the stress tensor is ambiguous. 
To address the issue, we will therefore consider other physical observables, namely correlation functions of $\zeta$ and $\gamma$. Since the couplings to $\zeta$ and $\gamma$ are gauge invariant, we expect these expectation values to carry physical information about the backreaction of the higher-spin fields.\label{foot:StressTensor}} We will return to this issue in~\S\ref{sec:tgreater1}. 

\subsection{Couplings to Inflationary Fluctuations}\label{sec:couplings}

When the inflationary fluctuations are coupled to PM fields, one faces the additional challenge that the allowed interactions must respect the PM gauge symmetry responsible for protecting the masses of these fields. We find that it is possible to write consistent couplings which generate an interaction of the form $\zeta\zeta\sigma$, as well as quadratic mixing terms $\gamma\sigma$, on the inflationary background. The consistency of these couplings should be viewed as an effective field theory (EFT) statement. The actions written below will likely require additional degrees of freedom to remain self-consistent at higher orders in interactions, i.e.~in order to maintain the gauge symmetry at each order in the fluctuations. Nonetheless, if such self-consistent theories exist, and are weakly coupled, then the main contribution to inflationary correlations should come from the leading, minimal vertices presented below.

\vskip 4pt
We write a minimal coupling of the form\footnote{We will only consider couplings to even-spin fields in this paper. For odd spins, the conserved currents need to involve at least two different fields, which can be achieved e.g.~with a complex scalar.}
\beq
	g_{\rm eff}\int\d^4x\sqrt{-g}\, \sigma_{\mu_1\cdots\mu_s}J^{\mu_1\cdots\mu_s}(\phi)\, ,\label{SigmaJSchematicCoupling}
\eeq
where $J^{\mu_1\cdots\mu_s}(\phi)$ is a spin-$s$ current, which depends quadratically on the inflaton field $\phi$, and $g_{\rm eff}$ is an effective coupling strength.\footnote{We will sometimes set $H=1$, so that $g_{\rm eff}$ becomes a dimensionless quantity.} This is analogous to the minimal coupling between the photon and charged matter in quantum electrodynamics (QED), $A_\mu J^\mu$.  As in QED, the gauge transformation of $\sigma_{\mu_1\cdots\mu_s}$ forces the fields contained in $J^{\mu_1\cdots\mu_s}$ to be charged under the PM gauge symmetry, i.e.$\!$ they must transform non-trivially if the action including \eqref{SigmaJSchematicCoupling} is to be gauge invariant off-shell. If we were to extend the theory to higher orders in the interactions, then the combined transformations would dictate the form of additional couplings needed for consistency, as well as possible deformations of the gauge transformations.  This is analogous to starting with a flat space coupling between a massless spin-2 field and a scalar, $h_{\mu\nu}\partial^{\mu}\phi\partial^{\nu}\phi$, demanding the linear gauge symmetry for $h_{\mu\nu}$ and, due to consistency, being led directly to the fully nonlinear, diffeomorphism-invariant theory of a  minimally-coupled scalar in general relativity (see e.g.~\cite{Deser:1969wk, Wald:1986bj}). Finding such a nonlinear completion of the PM theory is challenging, and we will content ourselves with studying the leading coupling presented above. For previous literature on the construction of the cubic couplings of PM fields, see e.g.~\cite{Joung:2012rv,Joung:2012hz}.

\vskip 4pt
The precise form of the current  $J^{\mu_1\cdots\mu_s}$ depends on the spin and the depth of the field $\sigma_{\mu_1\cdots\mu_s}$. For concreteness, we will study the example of a spin-$4$ field, but generalizations to higher spin are, in principle, straightforward.
 We will consider two special cases:  a massless spin-$4$ field and a partially massless spin-4 field with depth $t=1$. 
 
 \vskip 4pt
We first determine the conserved currents that can couple to these spin-4 fields. Under the gauge symmetry, the inflaton must transform in a non-trivial fashion.\footnote{Since the inflaton is charged under the gauge symmetry, it is possible for the inflaton background $\bar \phi(t)$ to affect the quadratic structure of the higher-spin action via higher-order couplings. To analyse this requires knowing the precise form of these couplings at quadratic order in the higher-spin field. In the present work, we assume that any such modifications are small.  We thank Paolo Creminelli and Andrei Khmelnitsky for discussions on this point.} Leaving the details to Appendix \ref{app:coupling}, we quote here the form of the current for a partially massless spin-4 field
\begin{align}
	J_{\mu\nu\rho\lambda}(\phi) = \nabla_{(\mu\nu}\phi\nabla_{\rho\lambda)}\phi + \cdots \, , \label{equ:J4}
\end{align}
  where the ellipses represent terms that vanish on-shell when contracted with $\sigma_{\mu_1\cdots\mu_4}$.  For the purpose of computing quantum expectation values, we expand the inflaton into the time-dependent background value and its perturbations, $\phi(t,\x)=\bar\phi(t)+\delta\phi(t,\x)$. In spatially flat gauge, (\ref{equ:J4}) and (\ref{SigmaJSchematicCoupling}) imply a cubic coupling between the field $\sigma_{\mu_1\cdots\mu_4}$ and $\delta \phi$. In terms of the gauge-invariant curvature perturbation, $\zeta = (H/\dot{\bar \phi}) \,\delta \phi$, this is
 \beq \label{vertexzetazetasigma}
	{\cal L}_{\sigma\zeta\zeta} \propto g_{\rm eff}\hs \dot{\bar\phi}^2\, \frac{\sigma_{ijkl} \partial_{ij} \zeta \hs \partial_{kl} \zeta}{a^8}\, ,
 \eeq
 where we have only kept the spatial components of the spin-4 field. While this is the unique non-vanishing term for a massless field, not all of the zero components vanish in the partially massless case. Although these non-spatial components can be included in our analysis, because they have the kinematical structure of lower-spin fields, we will opt for the full spatial components of the spinning field to show the characteristic spin-$s$ effect. Similar couplings exist for general spin-$s$, depth-$t$ fields, see Appendix~\ref{app:coupling}. 
The mixing between the graviton and a partially massless field can naturally be obtained by evaluating the coupling \eqref{SigmaJSchematicCoupling} on the inflaton background and perturbing the metric. The resulting coupling is
 \beq
	{\cal L}_{\gamma\sigma} \propto g_{\rm eff} \dot{\bar\phi}^{2} \, \frac{\sigma_{ij00}\dot\gamma_{ij}}{a^2}\, . \label{sigmagammafromg}
 \eeq 
The size of this mixing term is correlated with that of the cubic coupling. 

\vskip 4pt
To get an enhanced $\gamma \sigma$ coupling, it is desirable to write down an independent mixing operator.  
 For this purpose, we consider the following alternative coupling
\beq
	h_{\rm eff}\int\d^4x\sqrt{-g}\, \sigma_{\mu_1\cdots\mu_s}K^{\mu_1\cdots\mu_s}(\phi)\, ,\label{SigmaK}
\eeq
where $K^{\mu_1\cdots\mu_s}(\phi)$ is now {\it linearly} dependent on $\phi$ and $h_{\rm eff}$ is a coupling constant. Again, we can find the form of this coupling by demanding gauge invariance. The main difference between \eqref{SigmaJSchematicCoupling} and \eqref{SigmaK} is that, while we expect the former to be present in the standard minimal coupling scheme, the latter is an additional allowed coupling that we can introduce in our effective theory. For the case of spin 4, the current takes the form
 \begin{align}
 	K_{\mu\nu\rho\lambda}(\phi) = \nabla_{\mu\nu\rho\lambda}\phi+\cdots\, .
 \end{align}
Notice that this coupling vanishes when evaluated on the background metric. However, as shown in Appendix~\ref{app:coupling}, it still leads to a nontrivial $\gamma\sigma$ vertex of the form
 \beq
	{\cal L}_{\gamma\sigma} \propto h_{\rm eff} \dot{\bar\phi} \, \frac{\sigma_{ij00}\dot\gamma_{ij}}{a^2}\, .\label{sigmagamma}
 \eeq
Due to the number of zero components, these mixing terms are only non-trivial if the higher-spin field has depth $t\le 1$. The presence of $\dot{\bar\phi}$ in the prefactor indicates that mixing arises only when the conformal symmetry of the background is broken, consistent with the fact that $\langle\gamma\sigma\rangle$ vanishes identically when the conformal symmetry is exact.  

\vskip 10pt
In Section~\ref{sec:CC}, we will study the cosmological imprints of the interactions in (\ref{vertexzetazetasigma}), (\ref{sigmagammafromg}) and~(\ref{sigmagamma}).  To estimate the allowed sizes of the couplings $g_{\rm eff}$ and $h_{\rm eff}$, we write the interaction Lagrangian in terms of canonically normalized fields
\beq
{\cal L}_{\rm int} \,\sim\, g_{\rm eff} \,\frac{\sigma_{ijkl}\hskip 1pt \partial_{ij} \zeta_{c} \hskip 1pt\partial_{kl} \zeta_{c}}{a^8}  \,+\, \underbrace{\left(g_{\rm eff} \Delta_\zeta^{-1} + h_{\rm eff}\right)}_{\displaystyle \equiv \tilde h_{\rm eff}} \sqrt{r} \, \frac{\sigma_{ij00}\hskip 1pt \dot \gamma_{ij}^c}{a^2}\, .\label{SchematicInteractions}
\eeq
Here we are setting $H=1$, so that $ g_{\rm eff}$ and $h_{\rm eff}$ are dimensionless and that the fields $\zeta_{c} \sim \dot{\bar{\phi}}\,\zeta$ and $\gamma_{ij}^c \sim M_{\rm pl}\, \gamma_{ij}$ have order one fluctuations.
We have used $\dot{\bar{\phi}} \sim \Delta_\zeta^{-1}$ and $M_{\rm pl} \sim \Delta_\gamma^{-1}$, where $\Delta_\zeta$ and $\Delta_\gamma$ are the scalar and tensor amplitudes, respectively, and introduced the tensor-to-scalar ratio $r\equiv \Delta_\gamma^2/\Delta_\zeta^2$.  This rewriting of the Lagrangian makes manifest that the strength of the cubic interaction is determined by $g_{\rm eff}$ and the quadratic mixing by the combination $\tilde{h}_{\rm eff}\sqrt{r}$.

\vskip 4pt
The regime where \eqref{SchematicInteractions} remains perturbative depends on whether or not the interaction \eqref{SigmaK} is included in the action: 
\begin{itemize}
\item If \eqref{SigmaK} is not included, then $\tilde{h}_{\rm eff}=g_{\rm eff}\Delta_{\zeta}^{-1}$ and the quadratic and cubic couplings are both determined by $g_{\rm eff}$. In order for the quadratic mixing to remain weakly coupled, we then have to impose
\begin{align}
g_{\rm eff}\lesssim \frac{\Delta_{\zeta}}{\sqrt{r}}\, .\label{PerturbConditionsOnGeffOnly}
\end{align}
For $r > \Delta_{\zeta}^{2}$, this implies $g_{\rm eff} < 1$, ensuring that the cubic interaction is also perturbative.
\item If \eqref{SigmaK} is included in the action, then the quadratic mixing depends on a combination of $g_{\rm eff}$ and $h_{\rm eff}$, which we have denoted by $\tilde h_{\rm eff}$ in \eqref{SchematicInteractions}, while the cubic coupling is determined by $g_{\rm eff}$ alone.  In principle, this allows a large coupling to the tensor mode without a correspondingly large coupling to the scalars.\footnote{It would also be interesting to consider the case of strong mixing, $\tilde h_{\rm eff} > 1$, as a way to boost the tensor power spectrum without a corresponding effect on the scalar power spectrum. In this paper, we restrict ourselves to the more conservative case of weak mixing.}  
The requirements of weak mixing and perturbative interactions now place the following bounds on the couplings:\footnote{These bounds should be viewed as order-of-magnitude estimates, since various numerical factors appear in actual computations of correlation functions. The correct perturbativity conditions are expressed as bounds on the sizes of correlation functions, see Appendix~\ref{app:corr}.} 
\begin{align}
 \tilde h_{\rm eff}\lesssim \frac{1}{\sqrt{r}}\, , \quad g_{\rm eff}\lesssim 1\, . \quad\label{PerturbConditionsOnGeffAndHprimeeff}
\end{align}
For $r > \Delta_\zeta^2$, the couplings can only saturate \eqref{PerturbConditionsOnGeffAndHprimeeff} if the two contributions to $\tilde h_{\rm eff}$ cancel to a high degree. 
The net effect of including \eqref{SigmaK} in the action is to boost the possible size of both the tensor-scalar-scalar bispectrum and the scalar trispectrum; see \S\ref{sec:Bispectrum} and \S\ref{sec:Trispectrum}. 
\end{itemize}

\noindent
In~\cite{Lee:2016vti}, $\gamma\sigma$ interactions for massive spinning fields (i.e.$\!$ those belonging to the principal series) were considered in the context of the EFT of inflation~\cite{Cheung:2007st,Creminelli:2006xe}. In that case, it was found that the quadratic mixing $\gamma\sigma$  is always tied to the mixing $\zeta\sigma$, so that the weak coupling constraint of the latter induces a factor of $\sqrt r$ on the size of the graviton coupling, implying $\tilde h_{\rm eff}\lesssim 1$. In contrast, the fact that PM fields lack a longitudinal mode means that the mixing $\zeta\sigma$ vanishes on-shell. As a result, we expect the $\gamma\sigma$ vertex to remain weakly coupled under a less stringent condition as in \eqref{PerturbConditionsOnGeffAndHprimeeff}, not constrained by the size of the scalar coupling.

\section{Imprints on Cosmological Correlators}\label{sec:CC}

\begin{figure}[h!]
\centering
\includegraphics[height=100pt]{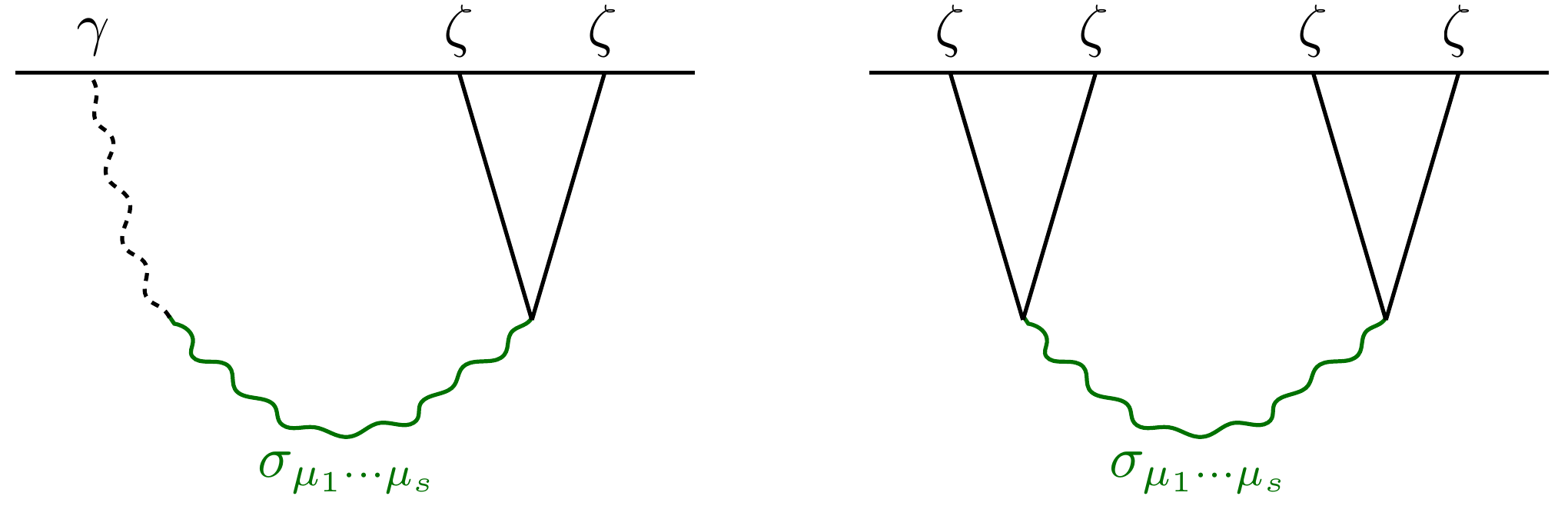}
\caption{\label{fig:corr} Tree-level diagrams contributing to $\langle\gamma\zeta\zeta\rangle$ (left) and $\langle\zeta\zeta\zeta\zeta\rangle$ (right).}
\end{figure}

We will now study the imprints of partially massless fields in cosmological correlators. In~\S\ref{sec:soft}, we briefly review the scalar and tensor consistency relations in single-field inflation. We then compute, in~\S\ref{sec:Bispectrum} and \S\ref{sec:Trispectrum}, the impact of an intermediate higher-spin field on both the tensor-scalar-scalar three-point function and the scalar four-point function (see Fig.~\ref{fig:corr}). We show that these fields can lead to a strict violation of the tensor consistency relation and a characteristic scaling and angular dependence in the four-point function, while giving no contribution to the scalar three-point function.

\subsection{Consistency Relations}
\label{sec:soft}

Symmetries play a crucial role in constraining inflationary correlation functions. In single-field inflation, $\zeta$ acts as the Goldstone boson of spontaneously broken conformal symmetries, nonlinearly realizing dilatations and special conformal transformations. The corresponding Ward identities imply consistency relations between correlation functions of different orders.

\vskip 4pt
For example, the Taylor expansion of the three-point function around the squeezed limit must take the form
\begin{align}
	\lim_{k_1\to 0}\langle\zeta_{\k_1}\zeta_{\k_2}\zeta_{\k_3}\rangle' = P_\zeta(k_1)P_\zeta(k_3)\sum_{n=0}^\infty a_n \left(\frac{k_1}{k_3}\right)^n\, , \label{zzz}
\end{align}
with $a_0 = 1-n_s$, where $n_s=0.968\pm 0.006$ is the scalar tilt~\cite{Ade:2015xua}. The leading term of the squeezed limit of the bispectrum is therefore fixed by the scale dependence of the power spectrum~\cite{Maldacena:2002vr}. The consistency relation also constrains $a_1$, and the model dependence in single-field inflation only enters at quadratic order \cite{Creminelli:2004yq,Creminelli:2012ed,Creminelli:2013cga,Pajer:2013ana, Hinterbichler:2013dpa}. This means that observing any non-analytic scaling behavior of the three-point function between $n=0$ and $n=2$ would be a clear signature of additional massive fields during inflation. Extra massless scalars would allow for a strict violation of the single-field consistency relation, i.e.~a modification of the coefficient $a_0$.

\vskip 4pt
Similar consistency relations exist in the tensor sector. For example, the tensor-scalar-scalar correlator can also be Taylor expanded around the squeezed limit
\begin{align}
	\lim_{k_1\to 0}\langle\gamma_{\k_1}^\lambda\zeta_{\k_2}\zeta_{\k_3}\rangle' = P_\gamma(k_1)P_\zeta(k_3)\,{\cal E}^\lambda_2(\hat\k_1\cdot\hat\k_3)\sum_{n=0}^\infty b_n\left(\frac{k_1}{k_3}\right)^n \, ,\label{gzz}
\end{align}
where ${\cal E}_s^\lambda(\hat\k_1\cdot\hat\k_3) \equiv \hat k^{i_1}_3\cdots \hat k_3^{i_s}\varepsilon^\lambda_{i_1\cdots i_s}(\hat \k_1)$ denotes the contraction between momenta and a spin-$s$, helicity-$\lambda$ polarization tensor.   
The two leading coefficients in this expansion are fully determined by symmetry; for example, $b_0=(4-n_s)/2$. This assumes that the only long-lived quadrupolar anisotropies are those sourced by the graviton. Again, there can be non-analytic contributions from extra massive particles as in the scalar case, but this time only with spin greater than or equal to two \cite{Lee:2016vti}. However, achieving a strict violation of the consistency relation for tensors turns out to be much more difficult. In particular, it was pointed out in~\cite{Bordin:2016ruc} that spin-2 particles cannot affect the leading term in the tensor consistency relation when the de Sitter symmetries are (approximately) respected, 
since unitarity forbids the existence of light spin-2 particles with mass $0<m^2<2H^2$. On the other hand, the particle spectrum allowed by dS representations is much richer and exists beyond spin two, as was reviewed in Section \ref{sec:HS}. It is then natural to ask whether higher-spin fields can affect this conclusion.

\vskip 4pt
Another interesting kinematical limit of cosmological correlators is the collapsed limit of the four-point function. This limit probes light states that are being exchanged in the four-point function, and it essentially factorizes into a product of three-point functions~\cite{Assassi:2012zq,Kehagias:2012pd,Kehagias:2012td,Pimentel:2013gza, Mirbabayi:2015hva}. This is analogous to the operator product expansion (OPE) limit of conformal correlation functions. In schematic form, we can write
\begin{align}
	\lim_{k_I\to 0}\langle\zeta_{\k_1}\zeta_{\k_2}\zeta_{\k_3}\zeta_{\k_4}\rangle' \, \sim\, \sum_\sigma \frac{\langle \zeta_{\k_1} \zeta_{-\k_1} \sigma_{\k_I}\rangle' \, \,\langle \sigma_{-\k_I} \zeta_{\k_3} \zeta_{-\k_3} \rangle'}{\langle\sigma_{\k_I}\sigma_{-\k_I}\rangle'}\, , \label{equ:col}
\end{align}
where $\k_I=\k_1+\k_2$ is an internal momentum. As in the OPE, the expression (\ref{equ:col}) involves a sum over all intermediate states, and the sum is dominated by the fluctuations that decay the slowest outside of the horizon.  See Appendix \ref{app:OPE} for more on this OPE perspective. 
\vskip 4pt
The Suyama-Yamaguchi (SY) relation~\cite{Suyama:2007bg} bounds the size of the collapsed trispectrum in terms of the size of the squeezed bispectrum~\cite{Smith:2011if, Assassi:2012zq}:
\beq
	\hat \tau_{\rm NL}\ge \left(\frac{6}{5} \hat f_{\rm NL}\right)^2\, ,\label{SY}
\eeq
where we have introduced the following nonlinearity parameters
\begin{align}
	\hat f_{\rm NL} \ &\equiv\ \lim_{k_1\to 0}\frac{5}{12}\frac{\langle\zeta_{\k_1}\zeta_{\k_2}\zeta_{\k_3}\rangle'}{P_\zeta(k_1)P_\zeta(k_3)}\, ,\label{fNL} \\
	\hat \tau_{\rm NL}\ &\equiv\ \lim_{k_{I}\to 0}\frac{1}{4}\frac{\langle\zeta_{\k_1}\zeta_{\k_2}\zeta_{\k_3}\zeta_{\k_4}\rangle'}{P_\zeta(k_1)P_\zeta(k_3)P_\zeta(k_{I})}\, ,\label{tNL}
\end{align}
and assumed that they have the same momentum scaling. The SY bound (\ref{SY}) is saturated when a single source is responsible for generating the curvature perturbations.\footnote{Strictly speaking, the SY bound is only saturated if the single source goes through a significant non-linear classical evolution on superhorizon scales. This is because, in single-field slow-roll inflation, the trispectrum from graviton exchange leads to a contribution $\hat\tau_{NL}={\cal O}(\varepsilon)$, which is in fact parametrically larger than $\hat f_{\rm NL}^2={\cal O}(\varepsilon^2)$~\cite{Seery:2008ax}. These slow-roll-suppressed effects are typically much smaller than the effects that we are interested in.} As we will show below, higher-spin fields provide an interesting example which nontrivially satisfies this bound.  In particular, we will show that PM fields do not generate any scalar bispectrum, while sourcing a nontrivial trispectrum. 

\subsection[Bispectrum: ${\langle \gamma\zeta\zeta\rangle}$]{Bispectrum: $\boldsymbol{\langle \gamma\zeta\zeta\rangle}$\label{sec:Bispectrum}}

We now study the effects of higher-spin particles on the $\langle\gamma\zeta\zeta\rangle$ correlator\footnote{Partially massless fields do not contribute to $\langle\zeta\zeta\zeta\rangle$ at tree level, as a spinning field without a longitudinal degree of freedom is kinematically forbidden to oscillate into a single scalar field.  Hence, $\langle \gamma \zeta\zeta\rangle$ is the simplest non-Gaussian correlator in which partially massless higher-spin fields can leave an imprint.} and demonstrate that they violate the consistency condition \eqref{gzz}. As was discussed in \S\ref{sec:couplings}, there exist two types of graviton couplings: one generated from the $\phi\phi\sigma$ coupling~\eqref{vertexzetazetasigma} and another from the $\phi\sigma$ coupling~\eqref{sigmagamma}. The former has the property that its signature in the tensor bispectrum is correlated with its effect on the scalar sector, while the latter allows the coefficients to be independently tuned, which can lead to an enhanced signal. 

\paragraph{Amplitude} We will use the following measure of the tensor-scalar-scalar bispectrum amplitude
\begin{align}
h_{\rm NL} \equiv \frac{6}{17} \sum_{\lambda=\pm2} \frac{\langle\gamma_{\k_1}^{\lambda}\zeta_{\k_2}^{\phantom\lambda}\zeta_{\k_3}^{\phantom\lambda}\rangle'}{P^{1/2}_\gamma(k)P_\zeta^{3/2}(k)}\, ,\label{fnltss}
\end{align}
where the bispectrum is evaluated in the equilateral configuration, $k_1=k_2=k_3 \equiv k$, with vectors maximally aligned with the polarization tensor. Our normalization agrees with \cite{Meerburg:2016ecv} and implies $h_{\rm NL}=\sqrt r/16$ for single-field slow-roll inflation \cite{Maldacena:2002vr}. 

\vskip 4pt
The size of the tensor bispectrum due to a higher-spin exchange can be estimated as 
\begin{align}
	h_{\rm NL} \sim \frac{\langle\gamma\zeta\zeta\rangle}{\langle\gamma^2\rangle^{1/2}\langle\zeta^2\rangle^{3/2}} 	\sim  g_{\rm eff}\sqrt{r}\Delta_\zeta^{-1}\times\begin{cases}g_{\rm eff}\Delta_{\zeta}^{-1}\,, \\[4pt] \tilde h_{\rm eff}\,,\end{cases}\label{PMhNL}
\end{align}
where the top and bottom cases correspond to excluding and including the  term \eqref{SigmaK} in the action, respectively. 
The perturbativity requirements  \eqref{PerturbConditionsOnGeffOnly} and \eqref{PerturbConditionsOnGeffAndHprimeeff} imply, respectively,
\begin{align}
	h_{\rm NL} \lesssim \begin{cases} \displaystyle r^{-1/2}\,, \\[4pt]  \Delta_{\zeta}^{-1}\,.\end{cases}\label{PMhNL}
\end{align}
The tensor bispectrum can be constrained using the $\langle BTT\rangle$, $\langle BTE\rangle$, $\langle BEE\rangle$ correlators of the CMB anisotropies~\cite{Meerburg:2016ecv, Abazajian:2016yjj}. The forecasted constraints from the CMB Stage IV experiments are $\sigma(\sqrt{r}h_{\rm NL})\sim 0.1$ for the local-type non-Gaussianity. The coupling $h_{\rm eff}$ is currently unconstrained, whereas the non-detection of the trispectrum puts an upper bound on $g_{\rm eff}$ (see \S\ref{sec:Trispectrum}) and consequently the size of the tensor bispectrum. 

\paragraph{Shape} As we described earlier, only higher-spin fields with depths $t=0$ or 1 contribute to $\langle\gamma\zeta\zeta\rangle$ at tree level. For $t=1$, the tensor bispectrum induced by the exchange of a PM field takes the following form in the soft limit 
\begin{align}
	\lim_{k_1\to 0}\frac{\langle\gamma^\lambda_{\k_1}\zeta_{\k_2}\zeta_{\k_3}\rangle'}{\alpha\Delta_{\raisemath{-0.5pt}{\gamma}}^{-1}} =  P_\gamma(k_1)P_\zeta(k_3)\, \hat Y^\lambda_s(\theta,\varphi)\, ,\label{TSSX}
\end{align} 
where $\hat Y^\lambda_s$ is a spherical harmonic, $\alpha$ is an effective coupling constant defined below, and the angles are defined by $\cos\theta=\hat\k_1\cdot\hat\k_3$ and $e^{i\varphi}=\eps\cdot\hat\k_3$. The polarization tensor for the graviton $\varepsilon_{ij}^\lambda$ is built out of two polarization vectors $\eps$, $\eps^*$ that span the plane perpendicular to $\hat \k_1$, which are fixed up to a phase (see~Appendix~\ref{app:A}). 
We note that the bispectrum \eqref{TSSX} has the same scaling as the leading term in the tensor consistency relation \eqref{gzz}.  Thus, fields with $t=1$, $s>2$ generate strict violations of \eqref{gzz} by moving the value of $b_{0}$ away from its predicted form. The case $t=1$, $s=2$  doesn't constitute a violation, as it corresponds to the massless graviton.  We dub this shape ``local tensor non-Gaussianity'', in analogy to the non-Gaussian shape that violates the scalar consistency relation. Due to the kinematics of the mixing with the graviton, only the helicity $\lambda=\pm 2$ modes of the higher-spin field contribute in the three-point function. Since these modes have the same amplitude, we have absorbed all of the numerical factors into $\alpha\sim g_{\rm eff} \tilde h_{\rm eff}\sqrt{r}$, which is required to be less than order unity in the weakly non-Gaussian regime. (The precise overall normalization  as a function of the spin of the field can be found in Appendix~\ref{app:corr}.) For $t=0$, the bispectrum is suppressed by a factor of $k_1/k_3$ relative to the $t=1$ case.

\begin{figure}[t!]
    \centering 
      \includegraphics[width=.5\textwidth]{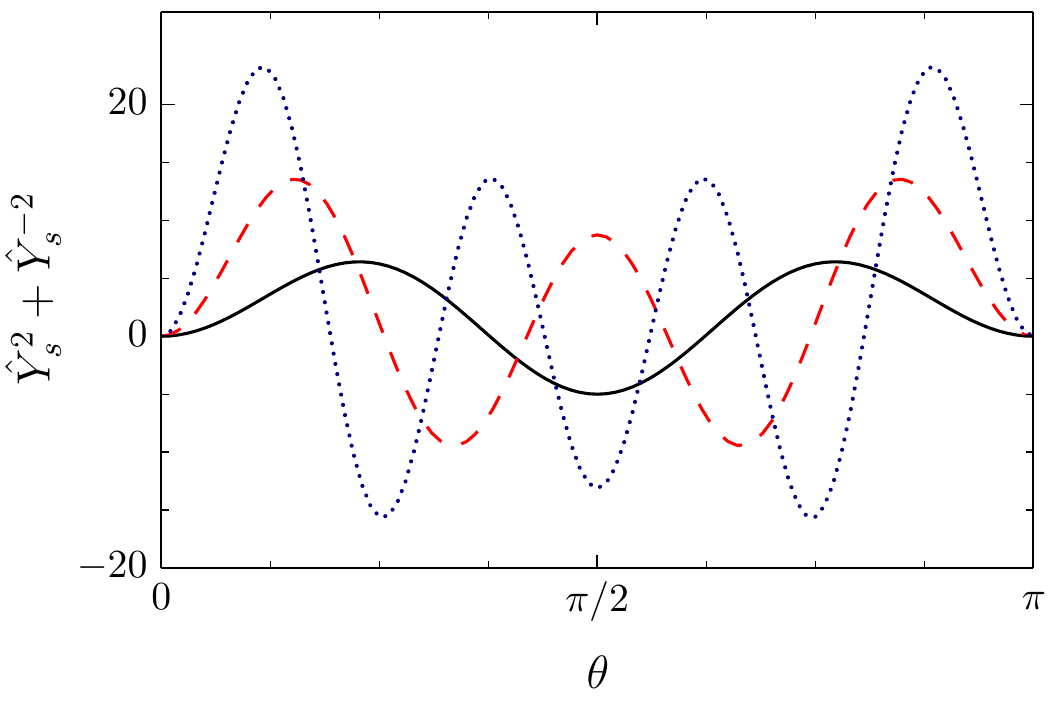}
    \caption{Angular dependence $\hat Y^2_s(\theta,0)+\hat Y^{-2}_s(\theta,0)$ due to the exchange of a depth-1 partially massless field as a function of the angle $\theta=\cos^{-1}(\hat\k_1\cdot\hat\k_3)$. The solid, dashed, and dotted lines correspond to spins 4, 6, and 8, respectively.}
    \label{fig:angle}
\end{figure}

\vskip 4pt
The angular dependence is given by the usual spherical harmonic of degree (spin) $s$ and order (helicity) $\lambda$, which can be factorized into longitudinal and transverse parts
\begin{align}
	\hat Y^\lambda_s(\theta,\varphi) = {\cal E}^\lambda_\lambda(\theta,\varphi) \hat P^\lambda_s(\cos\theta)\, ,
\end{align}
where $\hat P^\lambda_s(x)\propto (1-x^2)^{-\lambda/2}P^\lambda_s(x)$ is a version of the associated Legendre polynomial with a suitable normalization. 
On top of the usual quadrupole moment 
due to the external tensor mode, we see that the exchange of a higher-spin field induces an extra longitudinal angular component. 
Aligning $\k_1$ with the $z$-axis, we can express the hard momentum in spherical polar coordinates as $\hat\k_3=(\sin\theta\cos\varphi,\sin\theta\sin\varphi,\cos\theta)$. The angular dependence can then be written as a product of ${\cal E}^\lambda_\lambda=\sin^2\theta e^{\pm i\lambda\varphi}$ and the longitudinal part which is a function only of $\theta$. For example, some explicit expressions are
\begin{align}
	\sum_{\lambda=\pm 2}\hat Y_s^\lambda(\theta,\varphi) = \begin{cases}\displaystyle\frac{5}{2}(5+7\cos 2\theta)\sin^2\theta\cos 2\varphi & s=4 \\[10pt] \displaystyle\frac{36}{32}(35+60\cos 2\theta+33\cos 4\theta)\sin^2\theta\cos 2\varphi & s=6 \\[10pt] \displaystyle \frac{105}{256}(210+385\cos 2\theta+286\cos 4\theta+143\cos 6\theta)\sin^2\theta\cos 2\varphi & s=8\end{cases}\, .
\end{align}
Figure~\ref{fig:angle} shows the angular dependence as a function of the angle $\theta$ for $\varphi=0$. We can read off the spin of the particle by measuring the period of the oscillations. For the purpose of data analysis, having the full bispectrum shape available would also be helpful. The expression of the tensor bispectrum for general momentum configurations can be found in Appendix~\ref{app:corr}.

\vskip 4pt
Treating the soft tensor mode as a classical background, the squeezed tensor bispectrum would also contribute an anisotropic correction to the power spectrum in the following schematic way~\cite{Dai:2013kra}: 
\begin{align}
	\langle\zeta_{\k}\zeta_{\k'}\rangle_{\gamma_\q^\lambda}' = P_\zeta(k)\Big[1+h_{\rm NL}\hs \gamma_\q^\lambda \, \hat Y^\lambda_s(\theta,\varphi)\Big]\equiv P_\zeta(k)\Big[1+{\cal Q}_{i_1\cdots i_s}\hat k_{i_1}\cdots \hat k_{i_s}\Big]\, ,
\end{align}
for a given realization of the tensor mode, with the angles $\theta$ and $\varphi$ now defined in terms of $\k$ and $\q$. The external tensor mode is not directly observable, but its variance ${\cal Q}^2\equiv \langle {\cal Q}_{i_1\cdots i_s}^2\rangle$ is, after averaging over all long momenta. It would be interesting to measure this anisotropic effect due to PM fields in large-scale structure observations. See~e.g.~\cite{Jeong:2012df, Dimastrogiovanni:2014ina, Bordin:2016ruc} for forecasts on the detection limit of ${\cal Q}^2$ from future experiments and \cite{Bartolo:2017sbu} for related work on anisotropy generated by higher-spin fields.\footnote{When higher-spin fields acquire classical background values, they can also leave statistically anisotropic imprints in higher-order correlators~\cite{Franciolini:2017ktv}. We thank G.~Franciolini, A.~Kehagias, and A.~Riotto for sharing their draft with us.}

\subsection[Trispectrum: ${\langle \zeta\zeta\zeta\zeta\rangle}$]{Trispectrum: $\boldsymbol{\langle \zeta\zeta\zeta\zeta\rangle}$}\label{sec:Trispectrum}

We next consider the contribution of the interaction \eqref{vertexzetazetasigma} to the correlator $\langle\zeta\zeta\zeta\zeta\rangle$.  We will first focus on the special case of depth $t=1$ fields which freeze on superhorizon scales, and then study the imprints of $t>1$ fields which continue to grow after they exit the horizon. Fields of depth $t=0$ are not addressed in detail, as they decay on superhorizon scales and are hence phenomenologically less  interesting. Details of the relevant in-in calculations can be found in Appendix~\ref{app:corr}.

\subsubsection[{Depth $t=1$ Fields}]{Depth $\boldsymbol{t=1}$ Fields}

For a given spin $s\ge 2$, there exist $s-1$ partially massless states.  The depth $t=1$ case is rather special, as it is the only state which freezes on superhorizon scales.  In the following, we discuss the signature of such states on the $\langle \zeta\zeta\zeta\zeta\rangle$ correlator via the second diagram in Fig.~\ref{fig:corr}.

\paragraph{Amplitude}
We use the standard measure of the size of the trispectrum
\begin{align}
	\tau_{\rm NL}\ &\equiv\ \frac{25}{216}\frac{\langle\zeta_{\k_1}\zeta_{\k_2}\zeta_{\k_3}\zeta_{\k_4}\rangle'}{P^3_\zeta(k)}\, ,\label{tNL2}
\end{align}
where the right-hand side is evaluated at the tetrahedral configuration with $k_i=k$ and $\hat \k_i\cdot\hat \k_j=-1/3$. The estimated size of non-Gaussianity induced by the exchange of a PM field is
\begin{align}
	\tau_{\rm NL} \sim \frac{\langle\zeta^4\rangle}{\langle\zeta^2\rangle^3} \sim g_{\rm eff}^2\hs \Delta_\zeta^{-2}\, .
\end{align}
Imposing the weak coupling constraints \eqref{PerturbConditionsOnGeffOnly} and \eqref{PerturbConditionsOnGeffAndHprimeeff}, we find
\begin{align}
	\tau_{\rm NL} \lesssim \begin{cases} \displaystyle r^{-1}\,, \\[4pt]  \Delta_{\zeta}^{-2}\,,\end{cases}
\end{align}
where the top and bottom cases correspond to excluding and including the  term \eqref{SigmaK} in the action, respectively. Of course, the shape of the correlator crucially affects the observability of the signal. As we will describe below, the trispectrum under consideration has a similar scaling behavior in the soft limit as the ``local'' trispectrum, which arises in multi-field inflationary models (see \cite{Byrnes:2010em} for a review). For comparison, the current observational bound on the size of the latter is $\tau_{\rm NL}^{\rm local}=(-9.0\pm 7.7)\hs {\times}\hs 10^4$~\cite{Ade:2015ava}. This bound roughly translates into $g_{\rm eff}<10^{-2}$. Since there is no bispectrum counterpart for this signal, our scenario is an example which satisfies the SY relation in the most extreme manner.

\begin{figure}[t!]
\centering
\includegraphics[width=0.7\textwidth]{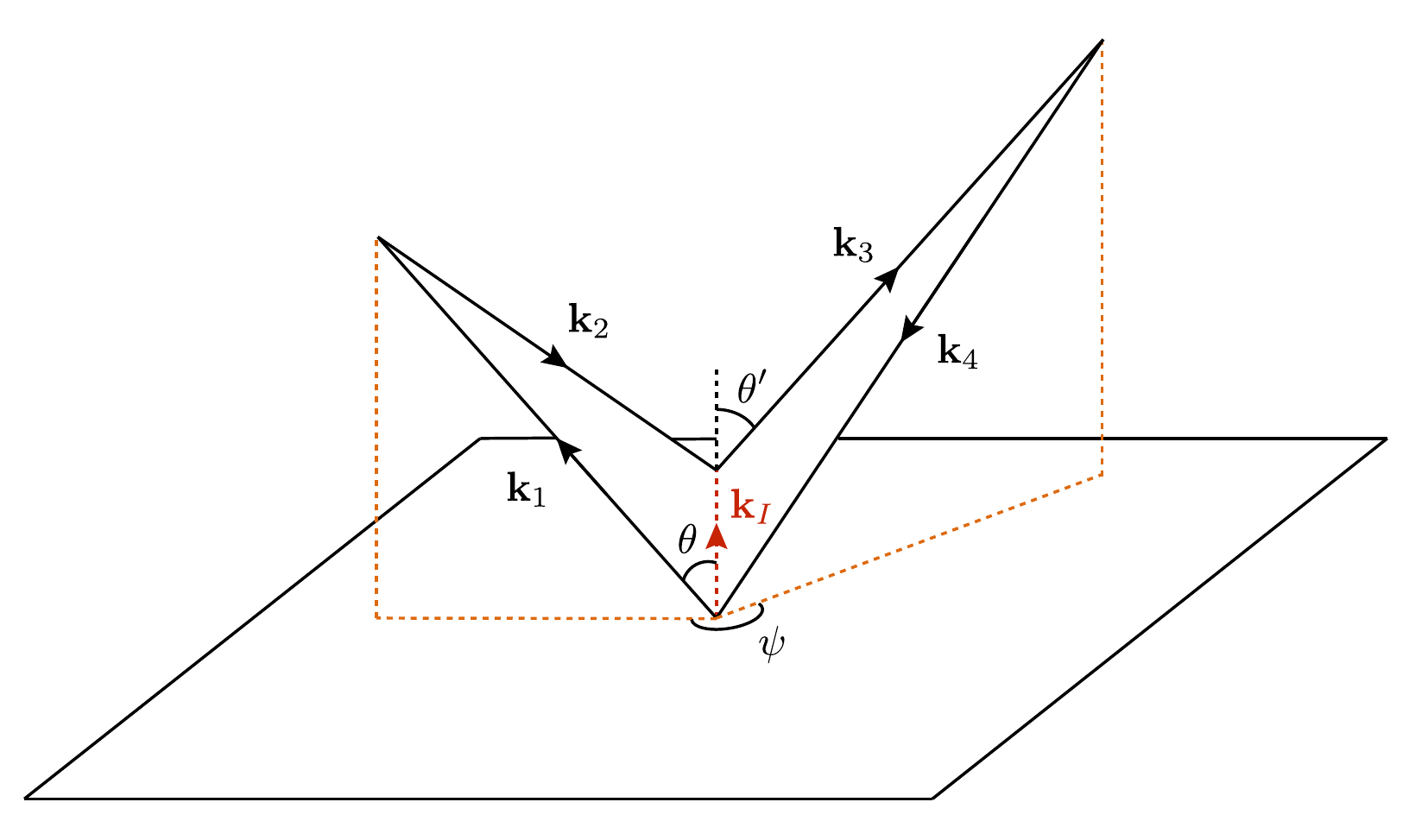}
\caption{\label{fig:tri} Sketch of the trispectrum in the collapsed configuration. The information on the spin of the exchanged (partially) massless field is contained in the $(\theta,\theta',\psi)$ dependence of the trispectrum.}
\end{figure}

\paragraph{Shape} Let us describe the collapsed limit of the trispectrum, leaving the details of the full shape to Appendix~\ref{app:corr}. For partially massless fields with depth $t=1$, we get 
\begin{align}
	\lim_{k_I\to 0}\frac{\langle\zeta_{\k_1}\zeta_{\k_2}\zeta_{\k_3}\zeta_{\k_4}\rangle'}{g_{\rm eff}^2\Delta_\zeta^{-2}}=P_\zeta(k_1)P_\zeta(k_3)P_\zeta(k_I)\sum_{|\lambda|=2}^s  E_s^{|\lambda|}\,\hat Y^\lambda_s(\theta,\varphi)\hat Y^{-\lambda}_s(\theta',\varphi')\, ,\label{spinscollapsed1}
\end{align}
where the amplitude of each helicity mode is 
\begin{align}
	E_s^\lambda = \frac{25[(2\lambda-1)!!]^2s!(s+1)!(s-\lambda)!(\lambda-2)!}{64\pi^2(2s-1)!!(s-2)!(s+\lambda)!(\lambda+1)!}\, . \label{Elambda1}
\end{align}
We see that the overall scaling behavior for a given depth is independent of the particle's spin. As advertised before, the scaling is the same as the local trispectrum, being proportional to $P_\zeta(k_1)P_\zeta(k_3)P_\zeta(k_I)$. The amplitude of each helicity mode is uniquely determined by its spin. The angular dependence is again factorized into the transverse part ${\cal E}^\lambda_\lambda$ and the longitudinal part $\hat P_s^\lambda$. The transverse polarization tensors project the momenta $\k_1$ and $\k_3$ onto the plane perpendicular to $\k_I$. The trispectrum is therefore a function of the angles $\cos\theta\equiv \hat\k_1\cdot\hat\k_I$ and $\cos\theta'\equiv \hat\k_3\cdot\hat\k_I$ between the vectors, as well as the angles $\varphi$ and $\varphi'$ on the projected plane with respect to the polarization tensor. The values of the projection angles depend on the two chosen polarization directions on the plane, but the difference $\psi\equiv \varphi-\varphi'$ is independent of this choice (see Fig.~\ref{fig:tri}). We can write  the full angular dependence as
\begin{align}
	\hat Y^\lambda_s(\theta,\varphi)\hat Y^{-\lambda}_s(\theta',\varphi') \propto e^{i\lambda\psi}P_s^\lambda(\cos\theta)P_s^{\lambda}(\cos\theta')  \, ,\label{YYProductBehavior}
\end{align}
where the factor of $(1-\cos^2\theta)^{\lambda/2}=\sin^{\lambda}\theta$ inside the associated Legendre polynomial contributes to the transverse part. Note that, as a consequence of the addition theorem for spherical harmonics, we have \cite{Arkani-Hamed:2015bza}
\begin{align}
	P_s(\cos\chi) \propto \sum_{\lambda=-s}^s \hat Y^\lambda_s(\theta,\varphi)\hat Y^{-\lambda}_s(\theta',\varphi')\, ,\quad 
\end{align}
where $\cos\chi = \cos\theta\cos\theta'+\sin\theta\sin\theta'\cos\psi$ is the angle between $\hat\k_1$ and $\hat\k_3$. This is not quite the angular dependence that we observe, since each helicity of the higher-spin field has a different amplitude, and some of the helicities are missing for the PM field. For $\{s,t\}=\{4,1\}$, we have
\begin{align}
	\sum_{|\lambda|=2}^4 E_4^{|\lambda|}\, \hat Y^\lambda_4(\theta,\varphi)  \hat Y^{-\lambda}_4(\theta',\varphi')\, \propto\,  \frac{15}{14}(5+7\cos 2\theta)(5+7\cos 2\theta')\sin^2\theta\sin^2\theta'\cos 2\psi &\nonumber\\[-5pt]
	\qquad + \, 75\cos\theta\cos\theta'\sin^3\theta\sin^3\theta'\cos 3\psi+\frac{105}{4}\sin^4\theta\sin^4\theta'\cos 4\psi &\, .\label{SumEYY}
\end{align}
As follows from \eqref{YYProductBehavior}, the helicity-$\lambda$ component of the field is responsible for the $\propto \cos\lambda\psi$ term in \eqref{SumEYY}.
We see that each helicity mode contributes with a distinct angular dependence, with all of the amplitudes being roughly of the same size. The factorization of the angular dependence into the sum over polarizations of the intermediate particle is analogous to what happens for flat-space scattering amplitudes when the intermediate particle goes on-shell, which is a consequence of unitarity. In the case of cosmological correlators, there is the precise relationship between the amplitudes of different helicities, which is a consequence of conformal symmetry.

\subsubsection[{Depth $t>1$ Fields}]{Depth $\boldsymbol{t>1}$ Fields}\label{sec:tgreater1}

We now study the impact of depth $t>1$ partially massless fields on the scalar trispectrum, which includes, in particular, the massless case $t=s-1$.  At tree level, this class of fields generates characteristic divergences in $\langle \zeta\zeta\zeta\zeta\rangle$, when evaluated in the collapsed configuration.

\vskip 4pt
In~\S\ref{sec:freeth}, we encountered the peculiar result that the two-point functions of depth $t>1$ partially massless fields diverge at late times; cf.~eq.~\eqref{sigma2pt}. 
However, the physical content of this fact is not clear, since the two-point function is gauge dependent.  The influence of the higher-spin field can instead be tested in a gauge-invariant manner by calculating its effect on $\langle \zeta\zeta\zeta\zeta\rangle$ via the exchange diagram in Fig.~\ref{fig:corr}.  In the collapsed limit, this trispectrum is directly sensitive to the higher-spin two-point function\footnote{This fact is made manifest in the wavefunction of the universe formalism; see Appendix \ref{app:OPE}.}
 and its scaling behavior due to the exchange of a spin-$s$, depth-$t$ PM particle is given by 
\begin{align}
	\lim_{k_I\to 0}\frac{\langle\zeta_{\k_1}\zeta_{\k_2}\zeta_{\k_3}\zeta_{\k_4}\rangle'}{g_{\rm eff}^2\Delta_\zeta^{-2}}=P_\zeta(k_1)P_\zeta(k_3)P_\zeta(k_I)\left(\frac{k_1k_3}{k_I^2}\right)^{t-1}\sum_{|\lambda|=2}^s  E_{s,t}^{|\lambda|}\,\hat Y^\lambda_s(\theta,\varphi)\hat Y^{-\lambda}_s(\theta',\varphi')\, ,\label{spinscollapsedfin}
\end{align}
with $E_{s,t}^\lambda$ defined in \eqref{Elambda}. Notice that this diverges at a rate that is faster than $P_\zeta(k_I)$ for $t>1$. 
\vskip 4pt
There are several effects which could make the strongly divergent behavior of \eqref{spinscollapsedfin}  less extreme.  First, loop diagrams with more intermediate higher-spin fields would also lead to a singular behavior that goes as $g_{\rm eff}^{2n}(k_1 k_3/k_I^2)^{n(t-1)}$, where $n$ is the number of loops.\footnote{This scaling behavior follows from a simple estimate of the contribution from ladder diagrams involving the cubic vertex at $n$ loops. For the contribution  due to vertices with more legs, a more detailed analysis is necessary.} Although these contributions are higher order in $g_{\rm eff}$, they would be more singular than the tree-level diagram if $t>1$. In that case, it does not make sense to only consider an individual diagram, but we would instead need to sum all of them.  It is conceivable that the result after the resummation will behave more tamely in the collapsed limit. Second, consistency of the theory may require the introduction of many new degrees of freedom and interactions which will also contribute to the scalar trispectrum.  If so, the multitude of particles may soften the collapsed limit of $\langle \zeta\zeta\zeta\zeta\rangle$.  This is the behavior claimed in \cite{Anninos:2017eib} which calculates, via a dual description, the four-point function of a conformally-coupled scalar due to the exchange of the tower of massless higher-spin fields in the minimal Vasiliev theory. In that case, the complete correlator behaves much more softly in the collapsed limit than the individual exchange diagrams do, which have similarly divergent behavior to \eqref{spinscollapsedfin}.

\section{Conclusions} \label{sec:Conc}
In this paper, we have studied the field theory of partially massless fields during inflation and discussed their imprints on cosmological correlators. Our main conclusions are:
\begin{itemize}
	\item Partially massless fields can have a consistent linearized coupling to a scalar field with arbitrary mass. We have constructed the corresponding conserved currents, and derived the relevant couplings between these higher-spin fields and the inflationary scalar and tensor perturbations. 
	\item Partially massless fields lead to a vanishing scalar bispectrum, but a non-zero trispectrum. The trispectrum has an unsuppressed behavior in the collapsed limit and a distinct angular dependence. 
	\item Partially massless fields can lead to a strict violation of the tensor consistency relation while respecting de Sitter symmetry, providing a loophole to the theorem of \cite{Bordin:2016ruc}. This local tensor non-Gaussianity is analogous to the sensitivity of the scalar bispectrum to extra light scalar species. 
	\item Partially massless fields can mix quadratically with tensor modes, but not scalar modes. The reason is purely kinematical---there is no longitudinal mode that mixes with the single scalar leg. This means that we can potentially enhance the tensor power spectrum, while not altering the scalar power spectrum. To realize this intriguing possibility would require understanding the regime of strong mixing of PM fields with the inflationary tensor modes.
	\item Partially massless fields do not decay outside the horizon, making them, in principle, directly observable after inflation.  
Predicting the final spectrum of PM fields is challenging, because it requires coupling these fields to the matter fluctuations in the late universe, something we did not address in this paper. It also requires understanding the meaning of PM fields away from the de Sitter limit.
\end{itemize}

\begin{figure}[t!]
    \centering
        \includegraphics[width=.9\textwidth]{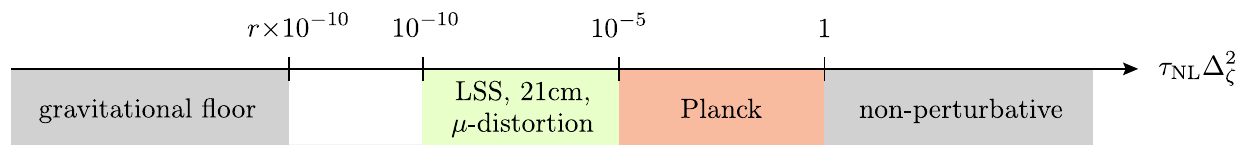}\\[4pt]
        \includegraphics[width=.9\textwidth]{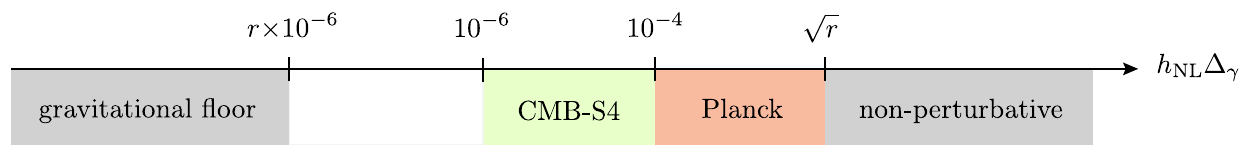}  
    \caption{Schematic illustration of current and future constraints on the scalar trispectrum (top)~\cite{Cooray:2008eb, Yamauchi:2015mja, Bartolo:2015fqz, Ade:2015ava} and tensor-scalar-scalar bispectrum (bottom)~\cite{Shiraishi:2010kd, Meerburg:2016ecv, Abazajian:2016yjj, Shiraishi:2017yrq} for local non-Gaussianity. The red and green regions correspond to the sensitivity levels of {Planck}~\cite{Ade:2015ava} and forthcoming experiments, respectively. The ``gravitational floor'' refers to the guaranteed level of non-Gaussianity sourced by gravitational nonlinearities during inflation, while ``non-perturbative'' denotes the strongly non-Gaussian regime. }     \label{fig:future}
\end{figure}

\noindent
Figure~\ref{fig:future} is a schematic illustration of current and future constraints on the scalar trispectrum and the tensor-scalar-scalar bispectrum. We see that future surveys will probe deeper into the allowed parameter space of primordial non-Gaussianities. 
We also see that there are still many orders of magnitude of parameter space left to be explored before we would hit the gravitational floor. Perhaps, within this unexplored territory, there will be new surprises and a rich cosmological fossil record waiting to be discovered. 

\paragraph{Acknowledgements} We thank Nima Arkani-Hamed, Matteo Biagetti, Cora Dvorkin, Kurt Hinterbichler, Austin Joyce, Andrei Khmelnitsky, Juan Maldacena, Soo-Jong Rey, David Vegh, and Matthew Walters for helpful discussions, and Dionysios Anninos, Paolo Creminelli and Antonio Riotto for comments on a draft. D.B., G.G.~and G.P.~acknowledge support from a Starting Grant of the European Research Council (ERC STG Grant 279617). G.G.~is also supported by the Delta-ITP consortium, a program of the Netherlands organization for scientific research (NWO) that is funded by the Dutch Ministry of Education, Culture and Science (OCW). H.L.~thanks the Institute for Advanced Study for hospitality while this work was completed. G.P.~acknowledges funding from the European Union's Horizon 2020 research and innovation programme under the Marie-Sk\l{}odowska Curie grant agreement number 751778. 
We acknowledge the use of the Mathematica package {\tt xAct}~\cite{xAct}.

\newpage
\appendix
\section{Higher-Spin Fields in de Sitter Space}
\label{app:A}
In this appendix, we describe the free theory of higher-spin fields in de Sitter space. In \S\ref{app:mode}, we derive the mode functions of partially massless higher-spin fields. Some general comments on PM fields and their relation to field theories with a global scaling symmetry are presented in~\S\ref{pmcomms}.

\paragraph{Preliminaries} It is useful to decompose quantum fields in terms of their helicities. We will work with the components of the spinning field $\sigma_{\mu_1\cdots\mu_s}$ projected onto spatial slices, i.e.~$\sigma_{i_1\cdots i_n \eta\cdots \eta}$.
We will find it convenient to write these components as
\beq
 \sigma_{i_1\cdots i_n \eta\cdots \eta} = \sum_\lambda\sigma_{n,s}^\lambda\hskip 1pt \varepsilon^\lambda_{i_1\cdots i_n}\, , \label{equ:A1}
\eeq 
where $\varepsilon^\lambda_{i_1\cdots i_n}$ is a suitably normalized totally symmetric spin-$s$, helicity-$\lambda$ polarization tensor, which satisfies the following properties:
\begin{align}
	\varepsilon^\lambda_{iii_3\cdots i_s}=0\, , \quad \hat k_{i_1}\cdots \hat k_{i_r}\varepsilon^\lambda_{i_1\cdots i_s}=0\, , \quad r> s-|\lambda|\, .\label{epsproperties}
\end{align}
There are three ``quantum number'' labels for the mode function $\sigma^\lambda_{n,s}$: the spin $s$, the ``spatial spin'' $n$, and the helicity $\lambda$ of the field. To avoid clutter, we will suppress the label for the depth $t$ of the field. Our convention for the polarization tensors is described in the insert below.
 
\begin{framed}
\small
\noindent{\it Polarization tensors}.---The polarization tensors acts as projection operators onto the angular momentum eigenfunctions. First, note that the properties \eqref{epsproperties} imply that the spin-$s$, helicity-$\lambda$ polarization tensor can always be decomposed as
\begin{align}
	\varepsilon_{i_1\cdots i_s}^\lambda(\hat\k,\eps) = \varepsilon_{(i_1\cdots i_\lambda}^\lambda(\eps) f_{i_{\lambda+1}\cdots i_s)}(\hat\k)\, ,\label{EpsilonTensor}
\end{align}
where $\varepsilon^\lambda_{i_1\cdots i_\lambda}$ is a maximally transverse polarization tensor which is constructed out of two polarization vectors $\eps^\pm$ that are perpendicular to $\hat\k$. The reality condition $\eps^+=(\eps^-)^*$ means that $\varepsilon^\lambda_{i_1\cdots i_\lambda}$ is just a function of a single polarization vector $\eps$, which can be specified up to a phase. The totally symmetric tensor $f_{i_1\cdots i_n}$ has the property that it becomes the longitudinal part of the associated Legendre polynomial upon contraction with momenta, i.e.~$\hat q_{i_1}\cdots \hat q_{i_n}f_{i_1\cdots i_{n}}(\hat\k)=\hat P^\lambda_s(\hat\q\cdot\hat\k)$ (see~\cite{Lee:2016vti}). The contraction of the polarization tensor with external momenta can therefore be expressed as
\begin{align}
	\hat q_{i_1}\cdots \hat q_{i_s}\varepsilon_{i_1\cdots i_s}^\lambda(\hat\k,\eps) = \hat Y_s^\lambda(\theta,\varphi) \, ,
\end{align}
where
\begin{align}
	\hat Y_s^\lambda(\theta,\varphi) & 	\propto e^{i\lambda\varphi}P^\lambda_s(\cos\theta)\equiv {\cal E}^\lambda_{\lambda}(\theta,\varphi)\hat P^\lambda_s(\cos\theta)\, ,
\end{align}
is a version of the usual spherical harmonics with a different normalization, with angles defined by $\cos\theta=\hat\q\cdot\hat\k$ and $\cos\varphi=\hat\q\cdot\eps$. We have denoted the transverse and longitudinal parts of the spherical harmonics as ${\cal E}^\lambda_\lambda$ and $\hat P^\lambda_s$, respectively. We will normalize the polarization tensors in such a way that the maximally transverse ($\lambda=s$) spherical harmonic has a unit coefficient:
\begin{align}
	\hat Y^s_s(\theta,\varphi) = e^{i\lambda\varphi}\sin^s\theta  \, .
\end{align}
This choice corresponds to
\begin{align}
	{\cal E}^\lambda_{\lambda}(\theta,\varphi) = e^{i\lambda\varphi}\sin^{\lambda}\theta \, , \quad \hat P^\lambda_s(\cos\theta) &= \frac{1}{(2\lambda-1)!!}\frac{\d^\lambda}{(\d \cos\theta)^\lambda}P_s(\cos\theta)\, ,
\end{align}
where it is left implicit that we are always taking the absolute value of $\lambda$ except in the phase $e^{i\lambda\varphi}$. In other words, we only distinguish opposite helicities by the phase, with $\hat P^{-\lambda}_s=\hat P^\lambda_s$. This normalization implies the following self-contraction of the polarization tensor
\begin{align}
	\varepsilon^\lambda_{i_1\cdots i_s}\varepsilon^{\lambda *}_{i_1\cdots i_s} = \frac{2^{s-\lambda}(2s-1)!!(s+\lambda)!}{[(2\lambda-1)!!]^2s!(s-\lambda)!}\, ,
\end{align}
with $\varepsilon^s_{i_1\cdots i_s}\varepsilon^{s *}_{i_1\cdots i_s}=2^s$.
\end{framed}

\subsection{Mode Functions}\label{app:mode}

A spin-$s$, depth-$t$ field satisfies the following on-shell equations of motion:
\begin{align}
	\Big[\Box-\big(s+2-t(t+1)\big)H^2\Big]\sigma_{\mu_1\cdots\mu_s}=0\, , \quad 	\nabla^{\mu}\sigma_{\mu\mu_3\cdots \mu_s}=0 \, , \quad {\sigma^{\mu}}_{\mu\mu_3\cdots \mu_s}=0\, .\label{PMeom}
\end{align}
These equations are invariant under the following gauge transformations~\cite{Hinterbichler:2016fgl}
\begin{align}
	\delta\sigma_{\mu_1\cdots\mu_s}	&=\begin{cases}\left[\prod_{n=1}^{u}\nabla_{\mu_n\mu_{n+u}}+(2n-1)^2H^2g_{\mu_n\mu_{n+u}}\right]_{\rm sym}\xi_{\mu_{s-t+1}\cdots\mu_s} & s-t\ \text{even} \\[5pt] \left[\prod_{n=1}^{w}\nabla_{\mu_n\mu_{n+w}}+(2n-1)^2H^2g_{\mu_n\mu_{n+w}}\right]_{\rm sym}\nabla_{\mu_{s-t}}\xi_{\mu_{s-t+1}\cdots\mu_s} & s-t\ \text{odd}
\end{cases}\, ,\label{PMgaugetrans}
\end{align}
where $[\cdots]_{\rm sym}$ denotes total symmetrization, $\{u,w\}\equiv \{\tfrac{s-t}{2},\tfrac{s-t-1}{2}\}$, and $\xi_{\mu_1\cdots\mu_t}$ is a totally symmetric gauge parameter that satisfies
\begin{align}
	\Big[\Box+\big((s-1)(s+2)-t\big)H^2\Big]\xi_{\mu_1\cdots\mu_t}=0\, , \quad \nabla^\mu\xi_{\mu\mu_2\cdots\mu_t}=0\, , \quad {\xi^\mu}_{\mu\mu_3\cdots\mu_t}=0\, .\label{PMgaugeonshell}
\end{align}
Unlike massless fields, PM fields consist of multiple helicity modes with $|\lambda|=t+1,\cdots, s$. To solve the on-shell equations~\eqref{PMeom}, we expand the field $\sigma_{\mu_1\cdots\mu_s}$ into its different helicity components. The helicity-$\lambda$ mode function with $n=|\lambda|$ polarization directions satisfies
\begin{align}
	{\sigma_{\lambda,s}^\lambda}''-\frac{2(1-\lambda)}{\eta}	{\sigma_{\lambda,s}^\lambda}'+\left(k^2-\frac{(t+\lambda-1)(t-\lambda+2)}{\eta^2}\right)\sigma_{\lambda,s}^\lambda = 0\, .
\end{align}
The solution with the Bunch-Davies initial condition is 
\begin{align}
	\sigma_{\lambda,s}^\lambda = Z_s^\lambda(-k\eta)^{3/2-\lambda}H_{1/2+t}^{(1)}(-k\eta)\, ,
\end{align}
where the normalization factor $Z_s^\lambda$ is given by
\begin{align}
(Z_s^{\lambda})^2 &= \frac{\pi}{4k}\frac{[(2\lambda-1)!!]^2s!(s-\lambda)!}{(2s-1)!!(s+\lambda)!}\frac{(\lambda+t)!(\lambda-t-1)!}{(s+t)!(s-t-1)!}\left(\frac{k}{H}\right)^{2s-2}\, .\label{Zs2}
\end{align}
This expression is well defined, since $t+1\le \lambda\le s$. Notice that the mode function involves the Hankel function of a half-integer index, which admits an expansion in terms of plane waves. Using the properties of the Bessel function, the mode function can also be expressed as
\begin{align}
	\sigma^\lambda_{\lambda,s}=Z_s^\lambda (-k\eta)^{3/2-\lambda}\hs e^{-ik\eta}\sum_{r=0}^t\frac{2i^{r-t-1}(t+r)!}{\sqrt{\pi}\hs r!(t-r)!(-2k\eta)^{r+1/2}}\, .
\end{align}
For the full angular dependence, we are interested in the spatial components of the field, i.e.~$\sigma^\lambda_{s,s}$. This can be obtained using the following recursive formula:
\begin{align}
\sigma_{n+1,s}^{\lambda} = -\frac{i}{k}\left({\sigma_{n,s}^{\lambda}}'-\frac{2}{\eta} \sigma_{n,s}^{\lambda}\right)-\sum_{m=|\lambda|}^n B_{m,n+1}\hskip 1pt \sigma_{m,s}^{\lambda}\, , \label{moderecur}
\end{align}
where the last term subtracts the trace part, with
\begin{align}
B_{m,n}\equiv \frac{2^n\hskip 1pt n!}{m!(n-m)!(2n-1)!!}\frac{\Gamma[\frac{1}{2}(1+m+n)]}{\Gamma[\frac{1}{2}(1+m-n)]}\, .
\end{align}
In other words, for a given helicity, the lowest `spatial spin' component determines the rest of the components. The derivation of the above formulas can be found in~\cite{Lee:2016vti}. As an example, let us provide the explicit expressions for the set of non-vanishing mode functions $\{\sigma^\lambda_{4,4}\}_{\lambda=2}^4$ for a $\{s,t\}=\{4,1\}$ field:
\begin{align}
	\sigma^4_{4,4}&= -i\,\frac{e^{-ik\eta}}{\sqrt{2k^3}}\frac{1+ik\eta}{4H^3\eta^4}\, ,\label{PMmode1}\\
	\sigma^3_{4,4}&= -\frac{e^{-ik\eta}}{\sqrt{2k^3}}\frac{5+5ik\eta-k^2\eta^2}{14\sqrt{5}H^3\eta^4}\, ,\label{PMmode2}\\
	\sigma^2_{4,4}&= i\,\frac{e^{-ik\eta}}{\sqrt{2k^3}}\frac{140+140ik\eta-55k^2\eta^2-13ik^3\eta^3}{70\sqrt{70}H^3\eta^4}\, .\label{PMmode3}
\end{align}
These mode functions all scale as $\eta^{-4}$ at late times. In contrast, the mode function of a massless spin-4 field is
\begin{align}
	\sigma^4_{4,4} = i\,\frac{e^{-ik\eta}}{\sqrt{2k^3}}\frac{15+15ik\eta-6k^2\eta^2-ik^3\eta^3}{4H^3k^2\eta^6}\, ,
\end{align}
which is more divergent than the PM field, scaling as $\eta^{-6}$ at late times.

\subsection{Comments on Partially Massless Fields}\label{pmcomms}

In this section, we make some general comments on PM fields, with the hope that these considerations might lead to a proper extension of PM fields beyond de Sitter space.

\vskip 4pt
Consider the action of a massless scalar field in flat space,
\beq
S=-\frac{1}{2}\int \d^4x\, (\nabla\phi)^2\, .
\eeq
This action has a shift symmetry $\phi \to \phi+c$ and a scaling (or Weyl) symmetry $x\to \Omega\hs x$, $\phi\to \Omega^{-2}\phi$. Gauging the Weyl symmetry leads to the action of a conformally-coupled scalar field \beq
S=-\frac{1}{2} \int \dd^4x\sqrt{-g}\left[ (\nabla\phi)^2+\frac{R}{6}\phi^2\right] .
\eeq
In de Sitter space, we have $R_{\mu\nu}= 3 H^2 g_{\mu\nu}$ and the mass of the field becomes $m^2=2H^2$. Thus, the confomally-coupled scalar field of $m^2=2H^2$ in de Sitter space has the same mode function as in flat space.  This is not all that surprising. What is surprising is that we can repeat this exercise for actions with higher derivatives, and something rather special happens in (A)dS.

\vskip 4pt
Now, consider the following action in flat space 
\beq\label{fda}
S=\kappa \int \d^4 x\, (\nabla^2 \phi)^2\,,
\eeq
which, once again, has a global scaling symmetry $x\to \Omega x$, $\phi \to \phi$. It also has an enhanced (Galileon-like) shift symmetry $\phi \to \phi + c_0 + c_1 \cdot x$. We can gauge the Weyl symmetry by coupling the field to a background metric and writing couplings to the curvature tensors. Around (A)dS, the resulting action can be rewritten as a sum of terms with four, two and zero derivatives. The terms with less derivatives carry more curvatures, to maintain the Weyl symmetry of the theory. Around the dS background, these are just factors of $H$. The surprising new feature is that the resulting action can be diagonalized. Performing a field redefinition of the schematic form $\xi\sim \nabla^2\phi$, the action can be written in two-derivative form, and then diagonalized. The normal modes will have $m^2 =0$ and $m^2=2H^2$. The massless field in dS is thus a descendant of the mode function satisfying the four-derivative action~\eqref{fda}. The coefficient $\alpha$ can be adjusted to set one of the normal modes to have positive kinetic term, but not both~\cite{Maldacena:2011mk}. This is a reflection of the Ostrogradsky instability of higher-derivative actions.

\vskip 4pt
To show this in more detail, we rewrite the action \eqref{fda} as follows
\beq\label{fsds}
S=\kappa \int \dd^4x\sqrt{-g} \left[(\nabla_g^2\phi)^2+2 (\nabla_g \phi)^2\right] ,
\eeq
where $g$ is the de Sitter metric and we have set $H = 1$ for convenience.\footnote{It is cumbersome to keep track of factors of $H$ in the intermediate manipulations, as $\phi$ has canonical dimension $0$ from the four-derivative action, and the diagonalized fields have dimension $1$. Once the diagonalization procedure is finished, it is trivial to restore the Hubble parameter.} This is just a reinterpretation of the same action. It can be viewed as a higher-derivative action either in de Sitter or in flat space. However, in de Sitter, we can do more with \eqref{fsds}. Introducing a Lagrange multiplier $\xi$ whose equation of motion is $\xi=\nabla_g^2\phi$, we can write the action in two-derivative form 
\beq
S=\kappa \int \dd^4x \sqrt{-g} \left[-2\nabla_g \xi\nabla_g\phi-\xi^2+2 (\nabla_g \phi)^2\right] .
\eeq
After a field redefinition $\xi-2\phi\equiv\varphi_1$ and $\xi\equiv \varphi_2$, the action can be rewritten as 
\beq
S=\frac{\kappa}{2}\int \dd^4 x \sqrt{-g} \left[(\nabla_g \varphi_1)^2- \left((\nabla_g \varphi_2)^2+2 \varphi_2^2\right) \right] ,
\eeq
which is the action for a conformally-coupled scalar field and a massless scalar field in de Sitter space. Thus, the mode functions of the original flat space action are also the mode functions of scalar fields in de Sitter space with $m^2=\{0,2H^2\}$. This explains the simplicity of the mode functions for this particular mass value $m^2=2H^2$, and also showcases the shift symmetry possessed by these actions. Notice that, regardless of the choice of $\kappa$, one of these fields will always have a wrong-sign kinetic term, thus implying a ghost instability of the original action. 

\vskip 4pt
We can repeat this exercise for higher-derivative actions. The next case would be a six-derivative action, which is diagonalizable around dS and has three normal modes, the previous two $m^2=\{0,2H^2\}$ and a third one of $m^2=-4H^2$. This explains why certain scalar tachyonic fields are special,\footnote{We thank David Vegh for pointing out the existence of these tachyonic modes to us.} in the sense of having divergent two-point functions: there is an additional shift symmetry that the spectrum possesses, and it is thus not possible to preserve de Sitter symmetry.
A nice way to see the role of the shift symmetries is to write the equations of motion for the scalar field as a conservation equation for the associated shift symmetry, $\nabla_\mu J^\mu=0$. It turns out that the shift symmetries have current densities associated to them. If we write the current as the density $J_p^\mu=\nabla^\mu (a^p \phi)$, where $a$ is the scale factor and $p$ is an integer related to the degree of the polynomial shift symmetry, then we can show that its conservation is the equation of motion for a scalar field in de Sitter of mass
\beq
m^2=-p(p-3) H^2\, ,
\eeq
which resembles the mass formula for operators of dimension $p$ in AdS/CFT.\footnote{The usual equation is $m^2=\Delta(\Delta-3)R_{\rm AdS}^{-2}$; it reduces to our equation under the analytic continuation $R_{\rm AdS}\to i H^{-1}$.} For example, the massless scalar in dS corresponds to the case $p=\{0,3\}$ and the conformally coupled scalar corresponds to the case $p=\{1,2\}$.

\vskip 4pt
As a side remark, we note that in AdS, the tachyonic scalar fields have positive masses and some higher shift symmetry. It would be interesting to see if they play any role in the AdS/CFT correspondence, and whether this shift symmetry protects their mass from quantum corrections.

\vskip 4pt
For the case of vector and tensor fields, a similar construction exists. Some care must be taken to ensure that the field excitations are transverse and traceless.\footnote{It is not clear that this can be ensured off-shell for arbitrary metrics in a Weyl-invariant fashion. We suspect that the higher-derivative actions can be coupled to any conformally flat background, although we haven't checked this claim. We thank Kurt Hinterbichler for emphasizing this point to us.} If we restrict ourselves to transverse and traceless fields, the action in flat space will schematically be of the form
\beq\label{highspac}
S \sim \int \dd^4x \left(\nabla^{n+1} h_s\right)^2\, .
\eeq
This action has a shift symmetry $\varepsilon\cdot h_s\to \varepsilon \cdot h_s+ c_0 + c_1 \cdot x+\cdots+c_n \cdot x^n $, where $\varepsilon$ is a suitable polarization tensor that projects $h_s$ onto its traceless transverse space. The action can be reinterpreted in any conformally flat background. Around dS, once again, the action diagonalizes in normal modes. These modes are parametrized by a depth label of the representation, and have mass
\beq
\frac{m^2}{H^2} = s(s-1)-t(t+1),~~t=0,1,\cdots, s-1\, .\label{MassOfSpinSDepthTParticle}
\eeq
 By increasing the depth, we obtain lighter fields, whose number of degrees of freedom is smaller than those of their heavier cousins. In more detail, the depth interpolates between the number of degrees of freedom of a massive field of spin $s$ in flat space; i.e., $2s+1$, and $2$, the number of degrees of freedom of a massless field.\footnote{In reality, the counting should begin with $2s+2$ and descend in pairs from the highest weight representation; the extra degree of freedom is ghost-like and decouples in the free theory 
\cite{Boulware:1973my}. In practice, the highest depth partially massless field has $2s$ degrees of freedom, and the reduction of its number of degrees of freedom by unity is attributed to Weyl symmetry.}  

\vskip 4pt
Finally, we should mention an interesting proposal to select the massless graviton action, using conformal gravity as a starting point \cite{Maldacena:2011mk} (see also \cite{Anastasiou:2016jix}). By imposing Dirichlet boundary conditions on the late-time amplitudes for the fluctuations,\footnote{These are the natural boundary conditions one must impose in order to compute the wavefunction of the universe. From the wavefunction of the universe one can extract inflationary correlation functions by using $|\Psi|^2$ as a probability distribution. When we compute inflationary expectation values directly, we do not impose late time boundary conditions on the field variables, but only that the initial state is the Bunch-Davies vacuum.} Maldacena showed that the wavefunction of the universe for Einstein gravity can be computed using conformal gravity. It would be interesting to study whether a similar procedure would select a partially massless field, using \eqref{highspac} as a starting point. The procedure outlined in \cite{Maldacena:2011mk} does not work for the case of the spin-2 field with $m^2=2H^2$ \cite{Deser:2012qg}, as ghost degrees of freedom appear in the interaction vertices beyond cubic order. Nonetheless, it was shown in~\cite{Deser:2012qg} that the spin-2 PM field can propagate in an Einstein spacetime (see also \cite{Bernard:2017tcg}). For our purposes, having a quadratic action for PM fields in a cosmological FRW background would suffice to determine whether the power spectra of higher-spin fields leave interesting imprints in the late universe. 
\newpage
\section{Spin-4 Couplings}\label{app:coupling}

In this appendix, we derive the form of the couplings of a scalar field and the graviton to a (partially) massless spin-4 field that were used in the main text. We also comment on the generalization of our couplings to partially massless fields of arbitrary spin.

\subsection{Conserved Current}\label{app:current}
We will construct the linearized coupling between (partially) massless spin-4 fields and a scalar field via the standard Noether procedure.\footnote{See \cite{Manvelyan:2004mb, Manvelyan:2009tf, Bekaert:2010hk} for similar Noether constructions of higher-spin currents for complex and conformal scalars in AdS.} For concreteness, we will present expressions for dS$_4$, but these results can easily be generalized to AdS or any number of dimensions.
We consider a coupling of the form
\begin{align}
	g_{\rm eff}\int\d^4x\sqrt{-g}\, \sigma_{\mu_1\cdots\mu_s}J^{\mu_1\cdots\mu_s}(\phi)\, .\label{sigmaJ}
\end{align}
First, we write the most general totally symmetric rank-$s$ tensor up to $s$ derivatives as
\begin{align}
	J_{\mu_1\cdots\mu_s}=\sum_{k=0}^{s/2} \beta_k\nabla_{(\mu_1\cdots \mu_k}\phi\nabla_{\mu_{k+1}\cdots\mu_s)}\phi + \cdots\, , \quad s=2,4,\cdots\, ,
\end{align}
where the ellipses denote terms that have contractions among the derivatives with appropriate factors of the metric. To begin with, let us construct a current for a massless spin-4 field. This field changes under a gauge transformation as
\begin{align}
	\delta\sigma_{\mu\nu\rho\sigma} = \nabla_{(\mu}\xi_{\nu\rho\lambda)}\, .\label{spin4gaugetrans}
\end{align}
Invariance of the coupling under this transformation implies that the current is conserved, $\nabla_\mu J^{\mu\nu\rho\lambda}=0$. We begin by writing down the most general form of a spin-4 current up to four derivatives:
\begin{align}
	J=\bigg[&\beta_1\phi\nabla^4\phi+\beta_2\nabla\phi\nabla^3\phi+\beta_3\nabla^2\phi\nabla^2\phi+\beta_4 g\nabla^2\phi\nabla^2\phi+\beta_5 g\nabla\phi\nabla^3\phi+\beta_6 H^2 g \nabla\phi\nabla\phi\nonumber\\
	&+\beta_7 H^2g\phi\nabla^2\phi+\beta_8 g^2\nabla^2\phi\nabla^2\phi+\beta_9 H^2 g^2 \nabla\phi\nabla\phi+\beta_{10} H^4 g^2\phi^2\bigg]_{\rm sym}\, ,\label{Jansatz}
\end{align}
where we have suppressed the indices and $g$ stands for the metric. Under the gauge transformation~\eqref{spin4gaugetrans}, the coupling $\sigma\cdot J$ will generate terms such as $ \nabla\xi\nabla^4\phi$. To cancel these terms, we also introduce a transformation rule for the scalar field
\begin{align}
	\delta\phi = (\lambda_1\nabla_{\mu\nu\rho}\phi+\lambda_2\nabla_\mu\phi\nabla_{\nu\rho}+\lambda_3\nabla_{\mu\nu}\phi\nabla_\rho+\lambda_4\nabla_{\mu\nu\rho})\xi^{\mu\nu\rho}\, .\label{spin4scalartrans}
\end{align}
We demand that the coupling is invariant under gauge transformations for both the spinning and scalar fields. We find that a massless spin-4 field can couple to scalar fields with arbitrary mass, at least at the linearized order.\footnote{Further imposing the tracelessness condition of the current forces the scalar to be conformally coupled with mass $m^2=2H^2$.}
After many integrations by parts and dropping boundary terms, the off-shell gauge invariance fixes the coefficients to be
\begin{align}
	\beta_3&=-3\beta_1+\tfrac{3}{2}\beta_2-\lambda_4\, , \quad \beta_4=4\beta_1-2\beta_2+\lambda_4\, , \quad \beta_5=-2\beta_1-\tfrac{1}{2}\beta_2\, , \quad \beta_6=26\beta_1-12\beta_2+4\lambda_4\, , \nonumber\\[3pt]
	\beta_7&=-26\beta_1\, , \quad \lambda_{2}=-3\beta_1+\tfrac{3}{4}\beta_2\, , \quad \lambda_{3}=-\beta_1\, , \quad \lambda_{4}=2\beta_1-\tfrac{1}{2}\beta_2+\lambda_3\, ,
\end{align}
where we have given the result for a massless scalar field. These conditions leave four parameters unconstrained, while we expect them to be fully fixed if the full theory were known. Nevertheless, this ambiguity does not have any observable consequences and goes away when we evaluate the coupling on-shell. In the transverse and traceless gauge, all terms in the current that are proportional to the metric drop out in the coupling. Essentially, the form of the coupling is uniquely fixed on-shell (up to terms that are related by integrations by parts). The current can be put in the form 
\begin{align}
	J_{\mu\nu\rho\lambda} = \nabla_{(\mu\nu}\phi\nabla_{\rho\lambda)}\phi\, .\label{Jspin4}
\end{align}
This leaves only one free parameter for the on-shell coupling, namely the coupling constant $g_{\rm eff}$. Moreover, all the zero components of the spin-4 field vanish on-shell.  
The covariant derivatives will lead to terms that involve $\Gamma^\mu_{ij} \propto \delta^\mu_0 \delta_{ij}$, but these do not contribute since the polarization tensor of $\sigma$ is traceless, $\varepsilon_{iijk}=0$. This means that we can simply replace the covariant derivatives in the current with partial derivatives. 
The on-shell current thus takes a particularly simple form
\begin{align}
	\sigma\cdot J = \frac{\sigma_{ijkl}\partial_{ij}\phi \hskip 1pt \partial_{kl}\phi}{a^8}\, .
\end{align}
One can perform a similar procedure of constructing an off-shell gauge-invariant linearized coupling between massless spinning fields and a scalar field for general spin. However, it is easy to see that the on-shell coupling will, again, take a unique form. Up to integration by parts, this is
\begin{align}
	\sigma\cdot J = \frac{\sigma_{i_1\cdots i_s}\partial_{i_1\cdots i_{s/2}}\phi \hskip 1pt \partial_{i_{s/2+1}\cdots i_{s}}\phi}{a^{2s}}\, .\label{GeneralExpectedSigmaJCoupling}
\end{align}
We will compute the correlation functions that arise from this coupling in Appendix~\ref{app:corr}.

\vskip 4pt
Next, let us consider the partially massless case. The on-shell gauge transformation of a spin-4, depth-1 partially massless field is [cf.~\eqref{PMgaugetrans}]
\begin{align}
	\delta\sigma_{\mu\nu\rho\sigma} =\nabla_{(\mu}\nabla_\nu\nabla_{\rho}\hs\xi_{\lambda)}+H^2g_{(\mu\nu}\nabla_\rho\xi_{\lambda)}\, ,\label{PMspin4gaugetrans}
\end{align}
with the gauge parameter subject to the conditions~\eqref{PMgaugeonshell}.\footnote{To ensure full off-shell gauge invariance, one needs to introduce a number of auxiliary lower-spin fields. Here we will restrict ourselves to the physical degrees of freedom of the on-shell PM field.} The coupling \eqref{sigmaJ} is invariant under this transformation if the current satisfies the condition
\begin{align}
	\nabla_{\mu\nu\rho} J^{\mu\nu\rho\lambda} + H^2 \nabla_\nu {J_{\mu}}^{\mu\nu\lambda}=0\, .
\end{align}
Again, we will start with the most general ansatz for the current \eqref{Jansatz}. Since the gauge transformation \eqref{PMspin4gaugetrans} contains more derivatives than that for massless fields, this time we require a scalar transformation involving five derivatives. It turns out that we can recast the most general totally symmetric scalar transformation in the form
\begin{align}
	\delta\phi \!=\! (\tilde\lambda_1 H^4\phi\nabla_\mu\! +\!\tilde\lambda_2 H^2\Box\nabla_\mu\phi\!+\!\tilde\lambda_3\nabla_{\mu\nu}\phi\nabla^\nu\!+\!\tilde\lambda_4\Box\nabla_{\mu\nu}\phi\nabla^\nu\!+\!\tilde\lambda_5\Box^2\nabla_\mu\phi\!+\!\tilde\lambda_6\nabla_{\mu\nu\rho}\phi\nabla^{\nu\rho})\xi^\mu\, ,\label{PMscalartrans}
\end{align}
via the use of the on-shell conditions for the gauge parameter $\xi^\mu$. The coupling will be invariant under the transformations \eqref{PMgaugetrans} and \eqref{PMscalartrans} if
\begin{align}
	\beta_3&=-\beta_1+\beta_2-\tilde\lambda_6\, , \quad \beta_5=\beta_4-\tfrac{1}{4}\tilde\lambda_3-2\tilde\lambda_4-2\tilde\lambda_6\, , \quad\nonumber\\
\tilde\lambda_1 &=-32\beta_1+8\beta_2-4\beta_4-4\beta_6+4\beta_7-3\tilde\lambda_2+3\tilde\lambda_3+15\tilde\lambda_4+36\tilde\lambda_6, , \quad \tilde\lambda_5=\tilde\lambda_4\, ,
\end{align}
for a massless scalar field. Again, the form of the coupling simplifies greatly on-shell, taking the form \eqref{Jspin4}. Evaluating the coupling in components, we can determine the coefficients of the couplings for both the spatial and non-spatial components of the PM field. Generalizing to higher spins and focusing on spatial components, the on-shell coupling, again, uniquely takes the form \eqref{GeneralExpectedSigmaJCoupling} for fields of any depth.

\subsection{Coupling to Gravity}
We are also interested in getting a $\gamma\sigma$ interaction vertex, which would contribute to the correlator $\langle\gamma\zeta\zeta\rangle$. This coupling can naturally be generated by evaluating the scalar fields in the coupling~(\ref{sigmaJ}) on the background and perturbing the metric.\footnote{The same procedure would also yield $\gamma\zeta\sigma$ and $\gamma\gamma\sigma$ interactions, which can contribute to the correlators $\langle\gamma\gamma\zeta\rangle$ and $\langle\gamma\gamma\gamma\rangle$, respectively.} In principle, this procedure could produce terms with and without derivatives acting on $\gamma_{ij}$. Naively, the latter would violate the tensor consistency relation even when $\sigma$ is massive, in contrast to the result of~\cite{Bordin:2016ruc}. However, we will show that such terms are indeed absent. 

\vskip 4pt
We take $\dot{\bar\phi}$ to be constant, which implies that $\nabla_{0\mu}\bar\phi=0$. Starting from \eqref{Jspin4}, we have
\begin{align}
	\sigma^{\mu\nu\rho\lambda}\nabla_{\mu\nu}\bar\phi\nabla_{\rho\lambda}\bar\phi=\sigma^{ijkl}\nabla_{ij}\bar\phi\nabla_{kl}\bar\phi &= \sigma^{ijkl}\hs \Gamma_{ij}^0\hs \Gamma_{kl}^0\hs \dot{\bar\phi}^2\nonumber \\
	&=a^{-4}H\Big[H\sigma_{iijj}-2H\sigma_{ijkk}\gamma_{ij}+\sigma_{ijkk}\dot\gamma_{ij}\Big]\dot{\bar\phi}^2\, ,
\end{align}
where the last expression follows from perturbing the metric in spatially flat gauge. We see that this also produces a tadpole term $\sigma_{iijj}$. Note that the spatial trace itself gives
\begin{align}
	g^{ij}g^{kl}\sigma_{ijkl} = a^{-4}(\sigma_{iijj}-2\sigma_{ijkk}\gamma_{ij})\, .
\end{align}
Demanding the absence of the tadpole exactly cancels the $\gamma_{ij}$ term. Consequently, the $\gamma\sigma$ vertex only involves $\dot\gamma_{ij}$. Using the traceless condition, we will denote this coupling for the general spin case by
\begin{align}
	g_{\rm eff}\dot{\bar\phi}^2\,\frac{\sigma_{ij0\cdots 0}\dot\gamma_{ij}}{a^{2s-2}}\, .\label{geffcoupling}
\end{align}
The resulting $\gamma\sigma$ vertex is proportional to $\dot{\bar\phi}^2$ and is correlated with the cubic vertex $\zeta\zeta\sigma$. 

\vskip 4pt
An alternative coupling between $\sigma$ and $\phi$ is of the form
\begin{align}
	h_{\rm eff}\int\d^4x\sqrt{-g}\, \sigma_{\mu_1\cdots\mu_s}K^{\mu_1\cdots\mu_s}(\phi)\, ,\label{sigmaK2}
\end{align}
where $K$ is linearly dependent on $\phi$.\footnote{Note that this coupling induces terms of order ${\cal O}(\phi\xi,\phi\sigma\xi)$ under the gauge transformations of $\sigma$ and $\phi$. The off-shell consistency of keeping both $\sigma\cdot J$ and $\sigma\cdot K$ types of couplings then requires introducing an extra scalar transformation rule of the form $\delta\phi = {\cal O}(\xi,\sigma\xi)$. This can always be done at linear order in $\sigma$.} The most general form of $K$ up to four derivatives is
\begin{align}
	K_{\mu\nu\rho\lambda} = \tilde\beta_1\nabla_{(\mu\nu\rho\lambda)}\phi+\tilde\beta_2g_{(\mu\nu}\nabla_{\rho\lambda)}\phi+\tilde\beta_3g_{(\mu\nu}\nabla_{\rho\lambda)}\Box\phi+\tilde\beta_4 g_{(\mu\nu}g_{\rho\lambda)}\Box\phi+\tilde\beta_5 g_{(\mu\nu}g_{\rho\lambda)}\Box^2\phi\, .\label{linearsigmaphi}
\end{align}
The last three terms vanish identically  upon using the equation of motion of $\phi$. Imposing the on-shell conservation condition $	\nabla_{\mu\nu\rho} K^{\mu\nu\rho\lambda} + H^2 \nabla_\nu {K_{\mu}}^{\mu\nu\lambda}=0$, we obtain
\begin{align}
	\tilde\beta_1=-16H^2\tilde\beta_2\, .
\end{align}
While the remaining terms also vanish when the background on-shell gauge conditions are imposed, $\bar g^{\alpha\mu}\nabla_{\alpha}\sigma_{\mu\nu\rho\lambda}=0$, $\bar g^{\mu\nu}\sigma_{\mu\nu\rho\lambda}=0$, they can still induce nontrivial couplings to the graviton. For example, we have
\begin{align}
	\sigma^{\mu\nu\rho\lambda}\nabla_{\mu\nu\rho\lambda}\bar\phi = \sigma^{ijkl}\nabla_{ijkl}\bar\phi &= \sigma^{ijkl}\Gamma^0_{ij}\Gamma^p_{0k}\Gamma^0_{pl}\hs \dot{\bar\phi}\nonumber\\
	&=a^{-4}H^2\Big[H\sigma_{iijj}-2H\sigma_{ijkk}\gamma_{ij}+\tfrac{3}{2}\sigma_{ijkk}\dot\gamma_{ij}\Big]\dot{\bar\phi}\, .
\end{align}
Again, only the $\dot\gamma_{ij}$ term remains after cancelling the tadpole. The other term gives
\begin{align}
	g^{\mu\nu}\sigma_{\mu\nu\rho\lambda}\nabla^{\rho\lambda}\bar\phi =g^{\mu\nu}\sigma_{\mu\nu ij}\nabla^{ij}\bar\phi&=g^{\mu\nu}g^{im}g^{jn}\sigma_{\mu\nu ij}\Gamma^0_{mn}\dot{\bar\phi}\nonumber\\
	&=a^{-4}H\Big[\tilde\sigma_{ii} -\gamma_{ij}\sigma_{ijkl}-\gamma_{ij}\tilde\sigma_{ij}+\tfrac{1}{2}\tilde\sigma_{ij}\dot\gamma_{ij}\Big]\dot{\bar\phi}\, ,
\end{align}
where $\tilde\sigma_{\mu\nu}=\bar g^{\alpha\beta}\sigma_{\alpha\beta\mu\nu}$ denotes the trace. On the other hand, the tadpole gives
\begin{align}
	g^{\mu\nu}g^{ij}\sigma_{\mu\nu ij}& =\bar g^{\mu\nu}\bar g^{ij}\sigma_{\mu\nu ij}+\delta g^{\mu\nu}\bar g^{ij}\sigma_{\mu\nu ij}+\bar g^{\mu\nu}\delta g^{ij}\sigma_{\mu\nu ij}\nonumber\\
	&=\tilde\sigma_{ii}-\gamma_{ij}\sigma_{ijkk}-\gamma_{ij}\tilde\sigma_{ij}\, .
\end{align}
The same logic prevents couplings involving $\gamma_{ij}$ without a time derivative.\footnote{The tadpole cancellation is more manifest in the language of the EFT of inflation, where graviton couplings are generated through the extrinsic curvature $\delta K_{ij} \supset\frac{1}{2}\dot{\gamma}_{ij}$ in comoving gauge.} We see that both terms will then lead to the generic form of the coupling
\begin{align}
	h_{\rm eff}\dot{\bar\phi}\,\frac{\sigma_{ij0\cdots 0}\dot\gamma_{ij}}{a^2}\, ,\label{heffcoupling}
\end{align}
which, in contrast to \eqref{geffcoupling}, is proportional to a single factor of $\dot{\bar\phi}$ and can be independent from the cubic vertex $\zeta\zeta\sigma$.

\newpage
\section{Cosmological Correlators}\label{app:corr}

In this appendix, we compute cosmological correlators involving an exchange of a (partially) massless field. We provide details of the computation of the scalar trispectrum and the tensor bispectrum in \S\ref{app:zzzz} and \S\ref{app:gzz}, respectively. In  Appendix~\ref{app:OPE}, we will analyze the soft limits of these correlators by applying the operator product expansion to both the wavefunction of the universe and to the final in-in correlator.

\paragraph{Preliminaries} The expectation value of an operator $ {\cal Q}$ at time $\eta_0$ is computed by 
\begin{align}
\langle\Omega| {\cal Q}(\eta_0)|\Omega\rangle = \langle 0| \left[\bar{\rm T} e^{i\int_{-\infty}^{\eta_0} \d \eta  \,H_I(\eta)}\right]{\cal Q}(\eta_0)\left[{\rm T} e^{-i\int_{-\infty}^{\eta_0} \d \eta \, H_I(\eta)}\right]|0\rangle\ ,\label{ininmaster}
\end{align}
where $|\Omega\rangle$ ($|0\rangle$) is the vacuum of the interacting (free) theory, $\rm T$ ($\bar{\rm T}$) denotes (anti-)time ordering, and $ H_I$ is the interaction Hamiltonian. We use the $i\epsilon$ prescription and replace $\eta\to\eta(1+i\epsilon)$ in the time integrals to evaluate the expectation value in the interacting vacuum. To compute quantum expectation values, we follow the usual procedure of quantization. We promote the fields $\zeta$ and $\sigma$ to operators and expand in Fourier space
\begin{align}
	\zeta(\eta,\k) = \zeta_k(\eta)a^\dagger(\k) +h.c. \, , \quad \sigma_{i_1\cdots i_n \eta\cdots \eta}(\k) = \sum_{|\lambda|=t+1}^s \varepsilon_{i_1\cdots i_n}^\lambda(\hat\k)\sigma_{s,s}^\lambda(k,\eta)b^\dagger(\k,\lambda) +h.c.\, ,\label{quantization}
\end{align}
where the creation and annihilation operators obey the canonical commutation relations
\begin{align}
	[a(\k),a^\dagger(\k')]=(2\pi)^3\delta(\k-\k')\, , \quad [b(\k,\lambda),b^\dagger(\k',\lambda')]=(2\pi)^3\delta_{\lambda\lambda'}\delta(\k-\k')\, .
\end{align}
The mode functions of $\zeta$ and $\gamma$ are given by
\begin{align}
	\zeta_k = \frac{H}{\dot{\bar\phi}}\frac{i}{\sqrt{2 k^3}}(1+ik\eta)e^{-ik\eta}\, ,\quad \gamma_k^\lambda = \frac{H}{\Mp}\frac{i}{\sqrt{2k^3}}(1+ik\eta)e^{-ik\eta} \, .
\end{align}
The mode functions of fields in the discrete series can be derived using the formulas given in Appendix~\ref{app:A}. The tree-level diagrams that we will compute have two interaction vertices. Expanding the in-in master formula \eqref{ininmaster} to quadratic order gives
\begin{align}
	\langle {\cal Q}(\eta_0)\rangle = \int_{-\infty}^{\eta_0}\d\eta\int_{-\infty}^{\eta_0}\d\tilde\eta \, \langle H_I(\eta){\cal Q}(\eta_0)H_I(\tilde\eta)\rangle-2{\rm Re}\int_{-\infty}^{\eta_0}\d\eta\int_{-\infty}^{\eta}\d\tilde\eta\, \langle {\cal Q}(\eta_0)H_I(\eta)H_I(\tilde\eta)\rangle\, ,
\end{align}
where it is understood that each side is evaluated in the appropriate vacuum state.

\subsection[Trispectrum: ${\langle \zeta\zeta\zeta\zeta\rangle}$]{Trispectrum: $\boldsymbol{\langle \zeta\zeta\zeta\zeta\rangle}$}\label{app:zzzz}
The result for the tree-level exchange of a spin-$s$ field is given by 
\begin{align}
\langle\zeta_{\k_1}&\zeta_{\k_2}\zeta_{\k_3}\zeta_{\k_4}\rangle' =  g_{\rm eff}^2\sum_{|\lambda|=t+1}^s\big({\cal T}_1-2{\rm Re}[{\cal T}_2]\big) + \text{23 perms}\, ,\label{AppCzeta4}\\
	{\cal T}_1 &=  \dot{\bar\phi}^4(k_1k_2k_3k_4)^{s/2}\zeta_1^*\zeta_2^*\zeta_3\zeta_4{\cal E}_{12I}^\lambda{\cal E}_{34I}^{-\lambda}\int_{-\infty}^0\frac{\d\eta}{a^{2s-4}}\, \zeta_1\zeta_2\sigma^\lambda_{I}\int_{-\infty}^0\frac{\d\tilde\eta}{a^{2s-4}}\,\zeta_3^*\zeta_4^*\sigma^{\lambda*}_{I} \, ,\\
	{\cal T}_2 &=  \dot{\bar\phi}^4(k_1k_2k_3k_4)^{s/2}\zeta_1\zeta_2\zeta_3\zeta_4{\cal E}_{12I}^\lambda{\cal E}_{34I}^{-\lambda}\int_{-\infty}^0\frac{\d\eta}{a^{2s-4}}\, \zeta_1^*\zeta_2^*\sigma^\lambda_{I}\int_{-\infty}^{\eta}\frac{\d\tilde\eta}{a^{2s-4}}\,\zeta_3^*\zeta_4^*\sigma^{\lambda*}_{I}\, ,
\end{align}
where we have suppressed the time arguments (e.g.~the mode functions outside of the integrals are evaluated at $\eta=0$) and defined 
\begin{align}
\sigma_i^\lambda &\equiv \sigma_{s,s}^\lambda(k_i)\, , \\
	{\cal E}_{abl}^\lambda &\equiv \hat k_a^{i_1}\cdots \hat k_a^{i_{s/2}}\hat k_b^{i_{s/2+1}}\cdots \hat k_b^{i_s}\varepsilon^\lambda_{i_1\cdots i_s}(\hat\k_l)\, .
\end{align}
We have also set $H=1$, which can be trivially restored. These integral formulas are even applicable for massive fields in the complementary and principle series upon setting $t=-1$ in \eqref{AppCzeta4}.

\paragraph{Spin-4 PM particles} Before presenting a formula for arbitrary spin, let us consider the example of a $\{s,t\}=\{4,1\}$ field. Using the mode functions~\eqref{PMmode1}--\eqref{PMmode3}, the non-time-ordered integral ${\cal T}_1$ can be expressed in the following compact form
\begin{align}
	{\cal T}_1 &= \sum_{|\lambda|=2}^4\I_1^{\lambda}\, , \\
	\I_1^\lambda &= (\pi\Delta_\zeta)^4 A_{|\lambda|}\,\frac{{\cal E}_{12I}^\lambda{\cal E}_{34I}^{-\lambda}}{k_1k_2k_3k_4k_I^3}\,\K_{|\lambda|}(k_1,k_2,k_I){\sf K}^*_{|\lambda|}(k_3,k_4,k_I)\, ,\label{Ilambda1}
\end{align}
where we have expressed the amplitude in terms of the dimensionless power spectrum and defined $A_2 ={1/32}$, $A_3 ={1/1960}$, $A_4 ={1/686000}$, and
\begin{align}
	\K_2(k_i,k_j,k_l)&=\J_0(k_i,k_j,k_l)+\J_1(k_i,k_j,k_l)\\
	\K_3(k_i,k_j,k_l)&=5\J_0(k_i,k_j,k_l)+5\J_1(k_i,k_j,k_l)+\J_2(k_i,k_j,k_l)\\
	\K_4(k_i,k_j,k_l)&=140\J_0(k_i,k_j,k_l)+140\J_1(k_i,k_j,k_l)+55\J_2(k_i,k_j,k_l)+13\J_3(k_i,k_j,k_l)\, ,
\end{align}
in terms of the function
\begin{align}
	\J_n(k_i,k_j,k_l)&\equiv -i\int_{-\infty}^0\d\eta\,(1+ik_i\eta)(1+ik_i\eta)(ik_l\eta)^ne^{-ik_{ijl}\eta}\nonumber\\
	&=\frac{k_l^n\big[n!\hs k_{ijl}^2+(n+1)!\hs k_{ij}k_{ijl}+(n+2)!\hs k_ik_j\big]}{k_{ijl}^{3+n}}\, ,\label{Jn}
\end{align}
with $k_{i_1\cdots i_s}\equiv k_{i_1}+\cdots+k_{i_s}$. The momentum scaling in \eqref{Ilambda1} is correct, since $\K_n$ (or $\J_n$) scales as $1/k$. For the time-ordered integral ${\cal T}_2$, the inner layer instead consists of an indefinite integral
\begin{align}
	&\L_n(k_i,k_j,k_l)=-i\int\d\eta\,(1+ik_i\eta)(1+ik_j\eta)(ik_l\eta)^ne^{-ik_{ijl}\eta}\nonumber\\
	&=\frac{k_l^n\big[k_{ijl}^2\Gamma(1+n,ik_{ijl}\eta)+k_{ij}k_{ijl}\Gamma(2+n,ik_{ijl}\eta)+k_ik_j\Gamma(3+n,ik_{ijl}\eta)\big]}{k_{ijl}^{3+n}}\, , \label{Ln}
\end{align}
where $\Gamma(n,x)$ is the incomplete gamma function, which takes the following form when $n$ is an integer
\begin{align}
	\Gamma(n,x)=(n-1)!\, e^{-x}\sum_{m=0}^{n-1}\frac{x^m}{m!}\, .
\end{align}
The function ${\cal T}_2$ can then be written as
\begin{align}
	{\cal T}_2 = \sum_{|\lambda|=2}^4\I_2^{\lambda}\, ,
\end{align}
where
\begin{align}
	\I_2^{\pm 4} & = \L_{00}+\L_{\{01\}}+\L_{11}\, ,\\[4pt]
	\I_2^{\pm 3} & = 25\hskip 1pt\L_{00}+25\hskip 1pt\L_{\{01\}}+5\hskip 1pt\L_{\{02\}}++25\hskip 1pt\L_{11}+5\hskip 1pt\L_{\{12\}}+\L_{22}\, ,\\[4pt]
	\I_2^{\pm 2} & = 19600\hskip 1pt\L_{00}+19600\hskip 1pt\L_{\{01\}}+7700\hskip 1pt\L_{\{02\}}+1820\hskip 1pt\L_{\{03\}}\nonumber\\
	&\hs +19600\hskip 1pt\L_{11}+7700\hskip 1pt\L_{\{12\}}+1820\hskip 1pt\L_{\{13\}}+3025\hskip 1pt\L_{22}+715\hskip 1pt\L_{\{23\}}+169\hskip 1pt\L_{33}\, ,\label{I24}
\end{align}
with $\L_{\{mn\}}\equiv\L_{mn}+\L_{mn}$ and
\begin{align}
	\L_{mn} &=(\pi\Delta_\zeta)^4\frac{{\cal E}_{12I}^\lambda{\cal E}_{34I}^{-\lambda}}{k_1k_2k_3k_4k_I^3}\int_{-\infty}^0\d\eta\,(1+ik_1\eta)(1+ik_2\eta)(ik_I\eta)^m\hs \L_n(k_3,k_4,k_I)\hs e^{i(k_I-k_1-k_2)}\, .
\end{align}
Although it is possible to obtain a closed-form expression for $\L_{mn} $, the result is lengthy and not very illuminating.

\vskip 4pt
In the collapsed limit, $k_I\ll k_1\approx k_2$, the expressions simplify dramatically. This is because the spin-4 mode function in the long-wavelength limit simplifies to
\begin{align}
	\sigma_{4,4}^\lambda (k\to 0)\propto \frac{e^{-ik\eta}}{H^3k^{3/2}\eta^4}\, .
\end{align}
In this case, the $k_I$ dependence can be pulled out of the integrals, after which it is easy to see that the trispectrum scales as $k_I^{-3}$ in the limit $k_I\to 0$. In this limit, we find
\begin{align}
	\I_1^{\lambda} + \I_2^\lambda &= 25(\pi\Delta_\zeta)^4C_{|\lambda|}\,\frac{{\cal E}_{11I}^\lambda{\cal E}_{33I}^{-\lambda}|_{s=4}}{(k_1k_3k_I)^3}\, ,
\end{align}
with coefficients $C_2 ={1/32}$, $C_3 ={5/392}$ and $C_4 ={1/35}$. Different helicity modes contribute with different but roughly similar amplitudes. The angular dependence now becomes
\begin{align}
	{\cal E}_{11I}^\lambda = {\cal E}^\lambda_s(\hat \k_1\cdot \hat\k_I) = {\cal E}^\lambda_\lambda(\hat \k_1\cdot \hat\k_I)P_s^\lambda(\hat \k_1\cdot \hat\k_I) \, ,
\end{align}
factorizing into the transverse and longitudinal parts. The final expression for the trispectrum in the collapsed limit is then
\begin{align}
	\lim_{k_I\to 0}\frac{\langle\zeta_{\k_1}\zeta_{\k_2}\zeta_{\k_3}\zeta_{\k_4}\rangle'_{s=4,t=1}}{g_{\rm eff}^2\Delta_\zeta^{-2}} = \frac{25}{8\pi^2}\hs P_\zeta(k_1)P_\zeta(k_3)P_\zeta(k_I)\sum_{|\lambda|=2}^4  C_{|\lambda|}\,\hat Y^\lambda_s(\theta,\varphi)\hat Y^\lambda_s(\theta',\varphi')\, ,
\end{align}
with angles defined in \S\ref{sec:Bispectrum}.

\paragraph{Arbitrary spin} The general formula for the exchange of a spin-$s$, depth-$t$ field in the collapsed limit is
\begin{align}
	\lim_{k_I\to 0}\frac{\langle\zeta_{\k_1}\zeta_{\k_2}\zeta_{\k_3}\zeta_{\k_4}\rangle'}{g_{\rm eff}^2\Delta_\zeta^{-2}} = P_\zeta(k_1)P_\zeta(k_3)P_\zeta(k_I)\left(\frac{k_1k_3}{k_I^2}\right)^{t-1} \sum_{|\lambda|=2}^s   F_t\, E_{s,t}^{|\lambda|}\, \hat Y^\lambda_s(\theta,\varphi)\hat Y^{-\lambda}_s(\theta',\varphi')\, ,\label{spinscollapsed}
\end{align}
where
\begin{align}
	E_{s,t}^{\lambda} \equiv \frac{s!(s+t)!(s-\lambda)!(\lambda-t-1)!}{(2s-1)!!(s+\lambda)!(\lambda+t)!(s-t-1)!}\left[\frac{(2\lambda-1)!!(2t)!}{2^{t+2}\pi\hs t!}\right]^2\, .\label{Elambda}
\end{align}
denotes the amplitude for each helicity and $F_t(k_1,k_3,k_I,\eta_0)$ is a function that can depend logarithmically on some of its arguments, with an IR cutoff $\eta_0$. The cases without any log dependences are $F_0 = {\cal O }(k_I^3)$, $F_1 = 25$ and $F_3 = 36$. Since this function is just constant for $t=1,3$, the scaling in $k_I$ for these cases is given by \eqref{spinscollapsed}. On the other hand, we see that there is an extra suppression for $t=0$ compared to the naive scaling that is suggested by the late-time behavior of the intermediate field. For general depths, the functional form of $F_t$ becomes more complicated as $t$ increases. However, the overall scaling behavior in $k_I$ is always fixed by $t$ as in \eqref{spinscollapsed} (with a few exceptions including $t=0$). The trispectrum therefore becomes singular in the collapsed limit for $t>1$, in the sense that it diverges at a rate faster than $P_\zeta(k_I)$ does. The derivation of \eqref{spinscollapsed} can be found in the insert below.

\begin{framed}
\small
\noindent {\it Derivation of~\eqref{spinscollapsed}.}---By taking the collapsed limit, we probe the late-time behavior of the intermediate field. The spin-$s$, helicity-$\lambda$ mode with spatial spin $n=\lambda$ behaves at late times as
\begin{align}
	\sigma^\lambda_{\lambda,s}(\eta\to 0)=Z_s^\lambda\hs  \frac{2^{1/2-t}(2t)!}{i\sqrt{\pi}\hs t!} \frac{e^{-ik\eta}}{(-k\eta)^{\lambda+t-1}}\, .\label{lambdalatetime}
\end{align}
The late-time behavior of the $n=\lambda+1$ mode can then be obtained using the recursion relation
\begin{align}
	\sigma^\lambda_{n+1,s}(\eta\to 0)= \frac{1+t+n}{-k\eta}\sigma^\lambda_{n,s}(\eta\to 0)\, .
\end{align}
This means that the $n=s$ mode will behave as
\begin{align}
	\sigma^\lambda_{s,s}(\eta\to 0)=Z_s^\lambda\hs  \frac{2^{1/2-t}(2t)!}{i\sqrt{\pi}\hs t!}\frac{(s+t)!}{(\lambda+t)!} \frac{e^{-ik\eta}}{(-k\eta)^{s+t-1}}\, . \label{lambdalatetime2}
\end{align}
It can then be shown that the two-point function in the late-time limit becomes
\begin{align}
	\sigma^\lambda_{s,s}\sigma^{\lambda*}_{s,s}(\eta\to 0) = \frac{8\pi^2 E_{s,t}^\lambda }{k^{1+2t}H^{2(s-1)}\eta^{2(s+t-1)}}\, ,
\end{align}
where the numerical prefactor given by \eqref{Elambda} fixes the amplitude of the helicity-$\lambda$ mode. In the small $k_I$ limit, the $k_I$ dependence drops out of the integrals, giving the overall scaling behavior as in \eqref{spinscollapsed}. The form of $F_t$ can be determined by computing the integral
\begin{align}
	(1-ik_1\eta_0)^2(1+ik_3\eta_0)^2e^{2i(k_1-k_3)\eta_0}
	\int_{-\infty}^{\eta_0} \frac{\d\eta}{\eta^{1-t}}\, (1+ik_1\eta)^2e^{-2ik_1\eta}\int_{-\infty}^{\eta_0} \frac{\d\tilde\eta}{\tilde\eta^{1-t}}\, (1-ik_3\tilde\eta)^2e^{2ik_3\tilde\eta} \, ,
\end{align}
and then taking $\eta_0\to 0$ limit and multiplying by an appropriate symmetry factor. 
\end{framed}

\subsection[Bispectrum: ${\langle \gamma\zeta\zeta\rangle}$]{Bispectrum:  $\boldsymbol{\langle \gamma\zeta\zeta\rangle}$}\label{app:gzz}
The tensor-scalar-scalar correlator with a general spin-$s$ field exchange is given by
\begin{align}
	\langle \gamma^\lambda_{\k_1}&\zeta_{\k_2} \zeta_{\k_3}\rangle' \equiv 2g_{\rm eff}h_{\rm eff}\sum_{\lambda'=\pm 2}{\rm Re}[{\cal B}_1-{\cal B}_2-{\cal B}_3]+\text{5 perms} \, ,\\
	{\cal B}_1&= \dot{\bar\phi}^3(k_1k_2)^{s/2}{\cal E}^{\lambda'}_{123}[\varepsilon_{ij}^{\lambda*}(\hat \k_3)\varepsilon_{ij}^{\lambda'*}(\hat \k_3)]\zeta_1^*\zeta_2^*\gamma_3^\lambda\int_{-\infty}^0\frac{\d\eta}{a^{2s-4}}\zeta_1\zeta_2\sigma_3^{\lambda'}\int_{-\infty}^0\frac{\d\tilde\eta}{a^{s-3}}(\gamma_3^{\lambda*})'\tilde\sigma_3^{\lambda'*}\, ,\label{B1}\\
	{\cal B}_2&=\dot{\bar\phi}^3(k_1k_2)^{s/2}{\cal E}^{\lambda'}_{123}[\varepsilon_{ij}^{\lambda*}(\hat \k_3)\varepsilon_{ij}^{\lambda'*}(\hat \k_3)]\zeta_1\zeta_2\gamma_3^\lambda\int_{-\infty}^0\frac{\d\eta}{a^{2s-4}}\zeta_1^*\zeta_2^*\sigma_3^{\lambda'}\int_{-\infty}^{\eta}\frac{\d\tilde\eta}{a^{s-3}}(\gamma_3^{\lambda*})'\tilde\sigma_3^{\lambda'*} \, ,\label{B2}\\
	{\cal B}_3&=\dot{\bar\phi}^3(k_1k_2)^{s/2}{\cal E}^{\lambda' *}_{123}[\varepsilon_{ij}^{\lambda*}(\hat \k_3)\varepsilon_{ij}^{\lambda'}(\hat \k_3)]\zeta_1\zeta_2\gamma_3^\lambda\int^0_{-\infty}\frac{\d\eta}{a^{s-3}}(\gamma_3^{\lambda*})'\tilde\sigma_3^{\lambda'}\int_{-\infty}^{\eta}\frac{\d\tilde\eta}{a^{2s-4}}\zeta_1^*\zeta_2^*\sigma_3^{\lambda' *}\, ,\label{B3}
\end{align}
where $\tilde\sigma_i^\lambda\equiv \sigma^\lambda_{2,s}(k_i)$ denotes the mode function with $n=2$ spatial components. These integral formulas are valid for fields with $t=-1,0,1$.\footnote{Recall that $t>1$ lacks a helicity-2 degree of freedom and that $t=-1$ indicates that the field belongs to either the complementary or principal series.} When summing over the PM field helicities, note that some of the integrals that involve taking the contraction of the transverse polarization tensors of the same helicity will vanish. This is because transverse polarization tensors are built out of two polarization vectors $\varepsilon_i^\pm$, which are null, $\varepsilon_i^\pm\varepsilon_i^\pm=0$, and satisfy $(\varepsilon_i^\pm)^*=\varepsilon_i^\mp$. This means that the only combination $\lambda=-\lambda'$ will contribute to ${\cal B}_1$ and ${\cal B}_2$, and $\lambda=\lambda'$ to ${\cal B}_3$. Our normalization for polarization tensors $\varepsilon^{s}_{i_1\cdots i_s}\varepsilon^{s *}_{i_1\cdots i_s}=2^s$ gives a factor of 4 from the contraction.

\paragraph{Spin-4 PM particles} Again, let us first specialize to the case $\{s,t\}=\{4,1\}$. The mode function for $n=\lambda=2$ is given by
\begin{align}
	\tilde\sigma_k^{\lambda=2} = -i\hs \frac{e^{-ik\eta}}{\sqrt{2k^3}}\frac{k^2(1+ik\eta)}{10\sqrt{70}\, \eta^2 H^3} \, .
\end{align}
Not surprisingly, this has exactly the same structure as the graviton mode function. The only difference is the spin-dependent normalization constant and extra powers of the scale factor due to the (conformal) time components of the PM field.

\vskip 4pt
With the mode functions being simple algebraic functions, it is lengthy but straightforward to compute the integrals \eqref{B1}-\eqref{B3} as in the previous section. Skipping the details of the computation, we find that the integrals can be expressed in the following compact form
\begin{align}
	{\cal B}_1 &=\frac{3\pi^3\Delta_\gamma^2\Delta_\zeta}{784000}\frac{{\cal E}^{\lambda' *}_{123}\delta_{\lambda\lambda'}}{k_1k_2k_3^3}\hs \K_4\, ,\\
	{\cal B}_2&=\frac{\pi^3\Delta_\gamma^2\Delta_\zeta}{784000}\frac{{\cal E}^{\lambda' *}_{123}\delta_{\lambda\lambda'}}{k_1k_2k_3^3}(-420\hskip 1pt\J_0^*+140\hskip 1pt\J^*_1-115\hskip 1pt\J^*_2+71\hskip 1pt\J^*_3-26\hskip 1pt\J^*_4)\, ,\\
	{\cal B}_3 &=\frac{\pi^3\Delta_\gamma^2\Delta_\zeta}{784000}\frac{{\cal E}^{\lambda' *}_{123}\delta_{\lambda\lambda'}}{k_1k_2k_3^3}(140\hskip 1pt{\sf P}_0+140\hskip 1pt{\sf P}_1+55\hskip 1pt{\sf P}_2+13\hskip 1pt{\sf P}_3)\, ,
\end{align}
where we have suppressed the arguments $(k_1,k_2,k_3)$ of the functions $\K_n$, $\J_n$, and ${\sf P}_n$, the latter of which is defined by
 \begin{align}
	&{\sf P}_n (k_i,k_j,k_l)  \equiv i \int (1+ik_l\eta)\L_n^*(k_i,k_j,k_l) =(n+1)! \, \frac{k_l^n}{2k_{ijl}^{5+n}}\nonumber\\
	&\times \Big\{ (n_5 k_i+k_{jl})(2n_3k_j^2+n_4k_l^2+n_4^2 k_jk_l )  +k_i^2\big[2n_3k_i+2(n_4^2-1)k_j+3(n_5^2-3)k_l\big]\Big\}\, ,
\end{align}
with $n_p\equiv n+p$. 

\vskip 4pt
When the external tensor mode becomes soft, $k_1\ll k_2\approx k_3$, a bunch of terms becomes unimportant and the bispectrum takes a considerably simpler form. First, we note that ${\sf P}_n$ scales as $k_1/k_3$ relative to $\K_n$ and $\J_n$ in the squeezed limit, which implies that we can neglect~${\cal B}_3$. Moreover, only $\J_0$ survives in the squeezed limit and $\J_{n>0}$ is subleading in $k_1$. Taking the permutations for which the PM field carries the soft momentum, the final result is given by
\begin{align}
	\lim_{k_1\to 0}\frac{\langle\gamma^\lambda_{\k_1}\zeta_{\k_2}\zeta_{\k_3}\rangle_{t=1}'}{\tilde\alpha\Delta_{\raisemath{-0.3pt}{\gamma}}^{-1}} =  P_\gamma(k_1)P_\zeta(k_3)\, \hat Y^\lambda_4(\theta,\varphi)\, ,\label{TSS}
\end{align} 
where $\tilde\alpha \equiv 3g_{\rm eff}\tilde h_{\rm eff}\sqrt{r}/2240\pi$ denotes an effective coupling strength, with angles defined in \S\ref{sec:Bispectrum}. Imposing the perturbativity bounds of \S\ref{sec:couplings}, this parameter naively needs to be much smaller than unity. However, the smallness of the overall numerical factor is due to the normalization of the $\sigma_{00ij}$ mode function in the $\gamma\sigma$ vertex, which should be taken into account when setting a bound on $\tilde h_{\rm eff}$. The correct perturbativity condition of the above correlator is then $\tilde\alpha\lesssim 1$.

\paragraph{Arbitrary spin} The tensor squeezed-limit bispectrum for an arbitrary spin-$s$ field with depth $t\in\{0,1\}$ is
\begin{align}
	\lim_{k_1\to 0}\frac{\langle\gamma^\lambda_{\k_1}\zeta_{\k_2}\zeta_{\k_3}\rangle'}{\alpha \Delta_{\raisemath{-0.3pt}{\gamma}}^{-1}} = P_\gamma(k_1)P_\zeta(k_3)\left(\frac{k_1}{k_3}\right)^{1-t}\hat Y^\lambda_s(\theta,\varphi) \, ,\label{TSSgeneral}
\end{align}
where we defined $\alpha \equiv N_{s,t}\, g_{\rm eff}\tilde h_{\rm eff}\sqrt{r}$, with
\begin{align}
	N_{s,t}= \frac{405\hs s!(s+t)!}{4^{t+2}\pi (2s-1)!!(s+1)!(s+2)!(t+2)!}\left[\frac{(2t)!}{t!}\right]^2\, .	\label{Nlambda}
\end{align}
Unlike in the case of the trispectrum, only the helicities $\lambda=\pm 2$ contribute. As a result, only $t\le 1$ fields can contribute, and the $t=0$ field leads to an extra suppression in the squeezed limit relative to the $t=1$ case. The derivation of the spin-dependent amplitude can be found in the insert below. This amplitude would be different for different types of $\gamma\sigma$ vertices. Typically, the higher the spatial spin of $\sigma$, the larger the amplitude. In other words, $N_{s,t}$ would be larger for interactions with more number of spatial derivatives, e.g.~$\sigma_{ijk\cdots}\partial_{k\cdots}\dot\gamma_{ij}$. Nonetheless, the weak coupling constraints imply $\alpha\lesssim 1$ for all types of interaction vertices.

\begin{framed}
\small
\noindent {\it Derivation of~\eqref{TSSgeneral}.}---Taking $\lambda=2$ in \eqref{lambdalatetime2}, we get
\begin{align}
	\sigma^{\lambda=2}_{s,s}(\eta\to 0)=Z_s^{\lambda=2} \frac{2^{1/2-t}(2t)!}{i\sqrt{\pi}\hs t!}\frac{(s+t)!}{(2+t)!} \frac{e^{-ik\eta}}{(-k\eta)^{s+t-1}}\, . 
\end{align}
while \eqref{lambdalatetime} gives
\begin{align}
	\sigma^{2}_{2,s}=Z_s^{\lambda=2} \frac{2^{1/2-t}(2t)!}{i\sqrt{\pi}\hs t!} \frac{e^{-ik\eta}}{(-k\eta)^{t+1}}
\end{align}
In the squeezed limit, the product of these mode functions becomes relevant
\begin{align}
	\sigma^{2}_{s,s}\sigma^{2*}_{2,s}(\eta\to 0) = \frac{4\pi \tilde N_{s,t} }{k^{2t-s+3}H^{2(s-1)}\eta^{s+2t}}\, ,
\end{align}
where the $\tilde N_{s,t}$ is related to \eqref{Nlambda} by some numerical factors. In the squeezed limit, ${\cal B}_1-{\cal B}_2$ becomes proportional to
\begin{align}
	\K_4(k_3,k_3,k_1\to 0)+140\hskip 1pt\J_0^*(k_3,k_3,k_1\to 0) = 280\hskip 1pt\J_0(k_3,k_3,k_1\to 0) = \frac{350}{k_1k_3}\, .
\end{align}
Also, the integration involving the $\gamma\sigma$ vertex gives a factor of $\frac{3}{4}$. Combining with other numerical and momentum-dependent factors in \eqref{B1} and \eqref{B2}, we arrive at \eqref{TSSgeneral}.

\end{framed}

\newpage
\section{Operator Product Expansions}\label{app:OPE}

In this appendix, we use the operator product expansion (OPE) to argue for the form of cosmological correlators in various limits.  
In particular, we will show how these arguments fix the form of the scalar trispectrum in the collapsed limit and the tensor-scalar-scalar bispectrum in the squeezed limit.

 \vskip 4pt
The OPE is used in two different ways in the derivation of the collapsed trispectrum.  First, it is applied directly to the collapsed limit of $\langle \zeta^{4}\rangle$.  Second, we assume that the coefficients in the  wavefunction of the universe have a good OPE and construct the collapsed limit of  $\langle \zeta^{4}\rangle$ given the restricted forms of these coefficients. Both methods yield the same result, but only when sufficient care is taken in deriving the OPE in momentum space.    

\vskip 4pt
We start, in \S\ref{wotun}, with a review of the wavefunction of the universe, and its connection to correlation functions in conformal field theory (CFT).  The behaviour of cosmological correlators in the soft limits can be recast in terms of the OPE in momentum space, which we consider in \S\ref{app:OPEinMomentumSpace}. We will address certain subtleties in performing the Fourier transform of more standard OPE expressions in position space. In~\S\ref{app:colltri}, we calculate the collapsed limit of $\langle \zeta^{4}\rangle$ using the two methods mentioned above.  Finally, the soft limit of $\langle \gamma \zeta\zeta\rangle$ is analyzed in \S\ref{app:softensbi}, using similar techniques.

\subsection{Wavefunction of the Universe}\label{wotun}

Consider a theory of a scalar $\zeta$ and a spin-$s$ field\footnote{Indices on $\sigma$, $\Sigma$ and related quantities will often be suppressed in this and following sections.} $\sigma$ on ${\rm dS}_{4}$.  The wavefunction for this system is determined semiclassically as $\Psi[\bar\zeta_{\k},\bar\sigma_{\k}]\approx \exp(iS_{\rm cl})$, where $S_{\rm cl}$ is the action evaluated on the classical solutions which interpolate from the Bunch-Davies vacuum at early times to the indicated value at late times: $\zeta_{\k}(\eta)\xrightarrow{\eta\to0}\bar{\zeta}_{\k}$ and $\sigma_{\k}(\eta)\xrightarrow{\eta\to 0}\bar{\sigma}_{\k}$. The classical action is then expanded as a function of the late-time field values
  \begin{align}
   \Psi[\bar{\zeta}_{\k},\bar{\sigma}_{\k}]&\approx \exp\bigg[\frac{1}{2}\int\d^{3}k_1\d^{3}k_{2}\,\big(\langle\mathcal{O}_{\k_{1}}\mathcal{O}_{\k_{2}}\rangle\bar{\zeta}_{\k_{1}}\bar{\zeta}_{\k_{2}}+\langle\Sigma_{\k_{1}}\Sigma_{\k_{2}}\rangle\bar{\sigma}_{\k_{1}}\bar{\sigma}_{\k_{2}}\big)\nn
   &\qquad  \ +\frac{1}{2}\int\d^{3}k_1\d^{3}k_{2}\d^{3}k_{3}\,\langle \mathcal{O}_{\k_{1}}\mathcal{O}_{\k_{2}}\Sigma_{\k_{3}}\rangle\bar{\zeta}_{\k_{1}}\bar{\zeta}_{\k_{2}}\bar{\sigma}_{\k_{3}}\nn
   &\qquad \ +\frac{1}{4!}\int\d^{3}k_1\d^{3}k_{2}\d^{3}k_{3}\d^{3}k_{4}\,\langle \mathcal{O}_{\k_{1}}\mathcal{O}_{\k_{2}}\mathcal{O}_{\k_{3}}\mathcal{O}_{\k_{4}}\rangle\bar{\zeta}_{\k_{1}}\bar{\zeta}_{\k_{2}}\bar{\zeta}_{\k_{3}}\bar{\zeta}_{\k_{4}}+\cdots \bigg]\, ,
   \end{align} 
   where the quantities in angled brackets are simply functions of the indicated momenta.
   Expectation values are then calculated by integrating the desired fields against $|\Psi[\bar{\zeta}_{\k},\bar{\sigma}_{\k}]|^{2}$, as in quantum mechanics.  For instance, this yields the late-time two-point functions
   \begin{align}
\langle \zeta_{\k}\zeta_{-\k}\rangle'&=-\frac{1}{2\hs \re \langle \mathcal{O}_{\k}\mathcal{O}_{-\k}\rangle'}\, , \quad \langle \sigma_{\k}\sigma_{-\k}\rangle'=-\frac{1}{2\hs\re \langle \Sigma_{\k}\Sigma_{-\k}\rangle'}\, ,\label{TwoPointFunctionsFromWU}
   \end{align}
   and the four-point function
\begin{align}
   &\langle \zeta_{\k_{1}}\zeta_{\k_{2}}\zeta_{\k_{3}}\zeta_{\k_{4}}\rangle'=  \frac{\langle{\cal O}^4\rangle'_A+\langle{\cal O}^4\rangle'_B}{\prod_{j=1}^{4}2\hs \re \langle{\cal O}_{\k_j}{\cal O}_{-\k_j}\rangle'}\, ,\label{FourPointFunctionFromWU}
\end{align}
with
\begin{align}
	\langle{\cal O}^4\rangle'_A &\equiv -\frac{2\hs\re\langle{\cal O}_{\k_1}{\cal O}_{\k_2}\Sigma_{-\k_{12}}\rangle'\, \re\langle\Sigma_{\k_{12}}{\cal O}_{\k_3}{\cal O}_{\k_4}\rangle'}{\re \langle\Sigma_{\k_{12}}\Sigma_{-\k_{12}}\rangle'}+\text{2 perms}\, , \label{O4A} \\[5pt]
	\langle{\cal O}^4\rangle'_B &\equiv 2\hs \re  \langle{\cal O}_{\k_1}{\cal O}_{\k_2}{\cal O}_{\k_3}{\cal O}_{\k_4}\rangle'\, ,
\end{align}
and $\k_{ij}\equiv \k_i+\k_j$.

\vskip 4pt
The wavefunction of the universe calculation is similar to AdS/CFT, a fact which we will return to below.
As in standard holographic computations, two conformal weights are associated to each bulk field:
\begin{align}
 \Delta_{\pm}&=\frac{3}{2}\pm\sqrt{ \frac{9}{4}-\frac{m^{2}}{H^{2}}}\ , \quad \Delta_{s\pm}=\frac{3}{2}\pm\sqrt{\left (s-\frac{1}{2}\right )^{2}-\frac{m^{2}}{H^{2}}}  \ ,\label{ConformalWeightsForFields}
\end{align}
for the scalar and spin-$s$ field, respectively.  In the limiting case where $\sigma$ is a partially massless, spin-$s$, depth-$t$ field, the weights $\Delta_{s\pm}$ reduce to
\begin{align}
\Delta_{s\pm}&=\frac{3}{2}\pm\left (\frac{1}{2}+t\right ) .\label{DeltaForSpin2DepthtField}
\end{align} The weights $\Delta_{+}$ and $\Delta_{s+}$ are assigned to the putative dual operators  $\mathcal{O}$ and $\Sigma$, whereas the bulk fields themselves are assigned the remaining weights,   $\Delta_{-}$ and $\Delta_{s-}$ for $\phi$ and $\sigma$, respectively.  More explicitly, the quantities in \eqref{DeltaForSpin2DepthtField} have the following scalings:
\begin{align}
\re\langle \mathcal{O}_{\k}\mathcal{O}_{-\k}\rangle'&\propto k^{2\Delta_{+}-3}\ \quad \Rightarrow  \quad \langle\zeta_{\k}\zeta_{-\k}\rangle'\propto k^{2\Delta_{-}-3}\, , \\
\re\langle \Sigma_{\k}\Sigma_{-\k}\rangle'&\propto k^{2\Delta_{s+}-3}\quad \Rightarrow \quad \langle\sigma_{\k}\sigma_{-\k}\rangle'\propto k^{2\Delta_{s-}-3}\ ,
\end{align}
where the relation $\Delta_{+}+\Delta_{-}=3$ was used, justifying the weight assignments.

\subsection{OPE in Momentum Space}\label{app:OPEinMomentumSpace}

In order to analyze the inflationary correlators in the kinematical regimes of interest, we must review some features of the OPE in momentum space. We begin with a discussion of the momentum space OPE between two scalar operators.  This is a subtle object,\footnote{We would like to thank Matthew Walters for discussions on this point.} as the process of Fourier transforming, in general, does not commute with taking the OPE limit~\cite{Dymarsky:2014zja,Bzowski:2014qja}.

 \vskip 4pt
As an illustrative example, consider the $\mathcal{O}^{(a)}\mathcal{O}^{(b)}\to \mathcal{O}^{(c)}$ position space OPE channel for three scalar operators with respective weights $\Delta_{a}$, $\Delta_{b}$ and  $\Delta_{c}$.  It is of the form
\begin{align}
\lim _{\x_{a}\to\x_{b}}\mathcal{O}^{(a)}(\x_{a}) \mathcal{O}^{(b)}(\x_{b})\propto \frac{1}{x_{ab}^{\Delta_{a}+\Delta_{b}-\Delta_{c}}}\, \mathcal{O}^{(c)}(\x_{b})\, ,\label{OPEAppendixPositionSpaceScalarOPE}
\end{align}
where $\x_{ab}=\x_{a}-\x_{b}$.  Fourier transforming both sides of \eqref{OPEAppendixPositionSpaceScalarOPE} gives
\begin{align}
\lim _{\q\to 0}\mathcal{O}^{(a)}_{\k-\q/2}\mathcal{O}^{(b)}_{-\k-\q/2}\propto  k^{\Delta_{a}+\Delta_{b}-\Delta_{c}-d}\, \mathcal{O}^{(c)}_\q\, ,\label{OPEAppendixNaiveScalarMomentumOPE}
\end{align}
in $d$-dimensions.  The result \eqref{OPEAppendixNaiveScalarMomentumOPE} is suspicious as it involves integrating an expression which only holds at close (but separated) points \eqref{OPEAppendixPositionSpaceScalarOPE} over all possible separations, and indeed \eqref{OPEAppendixNaiveScalarMomentumOPE} is not in general correct.

 \vskip 4pt
  The proper momentum-space OPE can instead be derived by directly Fourier transforming the explicit position space expression for the correlator $\langle \mathcal{O}^{(a)}(\x_{a})\mathcal{O}^{(b)}(\x_{b})\mathcal{O}^{(c)}(\x_{c})\rangle$.  The Fourier transform which isolates the desired momentum configuration is
\begin{align}
\langle \mathcal{O}^{(a) }_{\k-\q/2} \mathcal{O}^{(b) }_{-\k-\q/2} {\cal O}^{(c)}_{\raisemath{-1.3pt}{\q}} \rangle'
& = \int\d^{d}x\,\d^{d}y\, e^{i\k\cdot\x+i\q\cdot\y}\langle \mathcal{O}^{(a) }(\x/2) \mathcal{O}^{(b) }(-\x/2) \mathcal{O}^{(c) }(\y) \rangle\label{OPEAppendixSpecialOPEFourierTransform} \, ,
  \end{align}
and careful treatment \cite{Dymarsky:2014zja,Bzowski:2014qja} reveals distinct results depending on the weight $\Delta_{c}$. The correct form of this OPE channel is
\begin{align}
\lim _{\q\to 0} \mathcal{O}^{(a) }_{\k-\q/2} \mathcal{O}^{(b) }_{-\k-\q/2} \, \propto\, {\cal O}^{(c)}_{\raisemath{-1.3pt}{\q}}\, k^{\Delta_{a}+\Delta_{b}-\Delta_{c}-d}\times \begin{cases} 1& \Delta_{c}< {d/2}\\
\ln q & \Delta_{c}=d/2\\
({q/k})^{d-2\Delta_{c}} & \Delta_{c}> {d/2}
\end{cases}\, .\label{OPEAppendixScalarMomentumOPERightResult}
\end{align}
Hence, the naive result \eqref{OPEAppendixNaiveScalarMomentumOPE} is only correct in the regime of small weights, $\Delta_{c}\le {d/2}$.

\vskip 4pt
We will require  the same computation for the OPE channel from two scalars to a traceless, symmetric, spin-$s$ operator of weight $\Delta_{s}$: $\mathcal{O}^{(a)}\mathcal{O}^{(b)}\to \mathcal{O}_{i_{1}\cdots i_{s}}$. The end result is analogous to~\eqref{OPEAppendixScalarMomentumOPERightResult}.  The calculation requires the knowledge of the corresponding position space correlator, which is uniquely fixed by conformal invariance \cite{Costa:2011mg}:
\begin{align}
\langle\mathcal{O}^{(a)}(\x_{1})\mathcal{O}^{(b)}(\x_{2})\mathcal{O}_{s}(\x_{3})\rangle
&\propto \frac{\left (\z\cdot \x_{13}\, x_{23}^{2}-\z\cdot \x_{23}\,x_{13}^{2}\right )^{s}}{x_{12}^{(\Delta_{a}+\Delta_{b}-\Delta_{s}+s)/2}x_{13}^{(\Delta_{a}+\Delta_{s}-\Delta_{b}+s)/2}x_{23}^{(\Delta_{b}+\Delta_{s}-\Delta_{a}+s)/2}}\, .\label{OPEAppendixScalarScalarTensorPositionCorrelator}
\end{align}
Above,  we have contracted all loose indices of the spin-$s$ operator with an auxiliary null vector $z^{i}$ to turn it into the index-free operator $\mathcal{O}_{s}$. In order remove the auxiliary vectors, one repeatedly acts on the above with a particular derivative operator whose detailed form we will not  need. We then Fourier transform \eqref{OPEAppendixScalarScalarTensorPositionCorrelator} in the configuration \eqref{OPEAppendixSpecialOPEFourierTransform} following \cite{Dymarsky:2014zja}.  The process is straightforward and results in\footnote{Where the weights are such that the Fourier transform does not converge, these results are defined by analytic continuation.} 
\begin{align}
\lim _{\q\to 0} \mathcal{O}^{(a) }_{\k-\q/2} \mathcal{O}^{(b) }_{-\k-\q/2} \, \propto\,  \mathcal{O}_{i_{1}\cdots i_{s}}(\q)\, \hat k^{i_{1}}\cdots \hat k^{i_{s}} k^{\Delta_{a}+\Delta_{b}-\Delta_{s}-d}\times\begin{cases}
1 & \Delta_{s}< {d/2}\\
 \ln q & \Delta_{s}= {d/2}\\
 ({q/k})^{d-2\Delta_{s}}& \Delta_{s}>d/2
\end{cases}\ .\label{OPEAppendixGeneralOPEResult}
\end{align}  
In deriving \eqref{OPEAppendixGeneralOPEResult}, we have made use of the momentum space two-point function for spinning operators:
\begin{align}
\langle\mathcal{O}_{i_{1}\cdots i_{s}}(\q)\mathcal{O}_{j_{1}\cdots j_{s}}(-\q)\rangle'&\propto  q^{2\Delta_{s}-d}\, \pi_{i_{1}\cdots i_{s}j_{1}\cdots j_{s}}(\hat{\q})\, ,\label{OPEAppendixSpinningTwoPoint}
\end{align}
where $\pi_{i_{1}\cdots i_{s}j_{1}\cdots j_{s}}(\hat{\q})$ is a symmetric, traceless tensor structure.\footnote{The form of $\pi(\hat{\q})$ can be found using the arguments in Appendix A of \cite{Arkani-Hamed:2015bza}.  It is also related to the polarization vectors and normalization factors in Appendix \ref{app:A} via a completeness relation $\pi_{i_{1}\ldots i_{s}j_{1}\ldots j_{s}}\propto\sum_\lambda |Z_{s}^\lambda|^2 \varepsilon_{i_1\cdots i_s}^\lambda\varepsilon^{-\lambda}_{j_1\cdots j_s}$.} 
The result for $\Delta_{s}< {d/2}$ is the result which one would obtain from naively Fourier transforming the position space OPE of two scalar operators,\footnote{See \cite{Kehagias:2012pd}, for instance, for the form of this OPE.} while the cases of larger dimension are different, completely analogously to the scalar result \eqref{OPEAppendixScalarMomentumOPERightResult}.   The different behavior for large and small operator weights will be crucial for finding agreement between results derived via wavefunction of the universe and the in-in computations.

\subsection{Collapsed Trispectrum}\label{app:colltri}

We now use the OPE to derive the form of the collapsed limit of $\langle \zeta^{4}\rangle$ using the two methods discussed above.  We specialize to the case of massless $\zeta$ and take $\sigma$ to be a PM field of spin $s$ and depth $t$.

\vskip 4pt
First, we use the OPE to calculate the collapsed limit of $\langle \zeta^{4}\rangle$ directly.  The calculation is relatively straightforward.  The $\zeta$ and $\sigma$ fields have weight assignments $\Delta_{-}=0$ and $\Delta_{s-}=1-t$, respectively, and hence the $\Delta_{s}< {d/2}$ branch of the OPE applies, as $d=3$ here
\begin{align}
 \lim _{\q\to0}\zeta_{\k+ \q/2}\zeta_{-\k +\q/2}&=\sum_{s,t}C_{s,t}\,\sigma_{i_{1}\cdots i_{s}}(\q)\hat k^{i_{1}}\cdots \hat k^{i_{s}}k^{t-4}\label{OPENeededForInIn}\, ,
\end{align}
Performing the contractions and only keeping the contribution of a single spin-$s$, depth-$t$ operator, we find
 \begin{align}
 \lim _{\q\to 0}\langle\zeta_{\k_{1}- \q/2}\zeta_{-\k_{1}-\q/2}\zeta_{\k_{2}+ \q/2}\zeta_{-\k_{2}+ \q/2}\rangle' \propto \frac{k_{1}^{t-4}k_{2}^{t-4}}{q^{2t+1}}\left (\hat\k_{1}^{s}\cdot\pi(\hat{\q})\cdot \hat\k_{2}^{s}\right)\label{Collapsed4ptConfigurationResultInIn} ,
 \end{align}
 where $\pi(\hat{\q})$ is the tensor structure defined in \eqref{OPEAppendixSpinningTwoPoint}. This reproduces the scaling behavior in \eqref{spinscollapsed1}. The angular dependence can also be matched when the tensor structure is expanded into the helicity basis.

 \vskip 4pt
Next, we turn to the wavefunction of the universe.
Wavefunction coefficients arise as analytic continuations of AdS/CFT calculations, simply because in both cases one is computing the on-shell action as a function of boundary data \cite{Maldacena:2002vr,Anninos:2014lwa,Bordin:2016ruc}.   As the AdS quantities are CFT correlators which have a good OPE, it is feasible that the OPE can also be applied to wavefunction coefficients.  We now apply this logic to derive OPE limits of coefficients which in turn determine the collapsed limit of $\langle \zeta^{4}\rangle$.  The following arguments are intended to be more heuristic than rigorous and indeed we will find some technical disagreements between parts of the OPE prediction and concrete wavefunction calculations, though the discrepancies don't affect the predicted overall scaling of $\langle \zeta^{4}\rangle$.

 \vskip 4pt
 The four-point function is determined by $\langle\mathcal{O}^{4}\rangle'_A$ and $\langle\mathcal{O}^{4}\rangle'_B$, defined in \eqref{O4A}. In the collapsed configuration \eqref{Collapsed4ptConfigurationResultInIn}, the $\langle\mathcal{O}^{4}\rangle'_A$ term is dominated by just one of its three permutations, 
 \begin{align}
 \langle\mathcal{O}^{4}\rangle'_{A}\approx \lim _{\q\to 0}\frac{-2\hs\re\langle{\cal O}_{\k_1-\q/2}{\cal O}_{-\k_1-\q/2}\Sigma_{\q}\rangle'\, \re\langle\Sigma_{-\q}{\cal O}_{\k_{2}+\q/2}{\cal O}_{\k_{2}+\q/2}\rangle'}{\re \langle\Sigma_{\q}\Sigma_{-\q}\rangle'}\, ,
 \end{align}
 due to the strong scaling of the denominator with $q$, via the $\re \langle\Sigma_{\q}\Sigma_{-\q}\rangle'^{-1}$ factor. This is simply the PM power spectrum $\langle \sigma_{\q}\sigma_{-\q}\rangle' $, which is proportional to $q^{-1-2t}$. The terms in the numerator are evaluated using the OPE \eqref{OPEAppendixGeneralOPEResult} with $\Delta_{s}>d/2$, as opposed to \eqref{OPENeededForInIn}:
 \begin{align}
 \lim _{\q\to0}\mathcal{O}_{\k+ \q/2}\mathcal{O}_{-\k+ \q/2}&=\sum_{s,t}C_{s,t}\,\mathcal{O}_{i_{1}\cdots i_{s}}(\q)\hat k^{i_{1}}\cdots \hat k^{i_{s}}k^{t+2}q^{-2t-1}\label{OPENeededForWavefunction}\, ,
 \end{align} 
yielding
 \begin{align}
 \lim _{\q\to0}\re\langle{\cal O}_{\k-\q/2}{\cal O}_{-\k-\q/2}\Sigma_{\q}\rangle'&\propto k^{t+2}\, .
 \end{align}
Hence, in the collapsed limit, we find
 \begin{align}
\langle\mathcal{O}^{4}\rangle'_{A}\propto \frac{k_{1}^{t+2}k_{2}^{t+2}}{q^{2t+1}}\left ( \hat \k_{1}^{s}\cdot\pi(\hat{\q})\cdot \hat \k_{2}^{s}\right) ,\label{OPEResultForO4A}
 \end{align}
 where we restored the tensor structure.  The case of graviton exchange is calculated in \cite{Ghosh:2014kba} and the result is consistent with the scaling in \eqref{OPEResultForO4A}.  We have also calculated the result for massless spin-4 exchange and we again find agreement with \eqref{OPEResultForO4A}.

 \vskip 4pt
The OPE \eqref{OPENeededForWavefunction} can also be used to analyze the collapsed limit of $\langle\mathcal{O}^{4}\rangle'_{B}$, and here we find tension with concrete calculations. 
 Performing the contractions and keeping only the contribution from spin-$s$, depth-$t$ operators, one finds the same scaling as in  \eqref{OPEResultForO4A},
 \begin{align} 
\langle\mathcal{O}^{4}\rangle'_{B}\propto \frac{k_{1}^{t+2}k_{2}^{t+2}}{q^{2t+1}}\left (\hat \k_{1}^{s}\cdot \pi(\hat{\q})\cdot \hat \k_{2}^{s}\right) . \label{OPEResultForO4B}
 \end{align}
 However, unlike \eqref{OPEResultForO4A}, the result \eqref{OPEResultForO4B} is \textit{not} in agreement with the explicit calculations for massless spin-2\hskip 1pt\footnote{See \cite{Ghosh:2014kba} for more on the subdominance of  $\langle\mathcal{O}^{4}\rangle'_B$ to $\langle\mathcal{O}^{4}\rangle'_A$ in the spin-2 case.} or spin-4 particles, where this contribution is instead found to be subleading as $q$ is taken soft. 
  While this finding doesn't affect the final scaling for $\langle \zeta^{4}\rangle$, it does imply that the OPE cannot be naively applied\footnote{We can make an interesting speculation for the origin of the mismatch: in the putative dual theory, there may be more than one operator associated to $\sigma$.  In particular, if there is not only the operator $\Sigma$ of weight $\Delta_{s+}$, but also the associated \textit{shadow} operator $\tilde{\Sigma}$ of weight $\Delta_{s-}$ in the spectrum, then it's possible that the $\mathcal{O}\mathcal{O}\to \Sigma$ and $\mathcal{O}\mathcal{O}\to \tilde{\Sigma}$ OPE channel contributions to $\langle\mathcal{O}^{4}\rangle'_{B}$ can cancel out, while the existence of the shadow operator $\tilde{\Sigma}$ leaves $\langle\mathcal{O}^{4}\rangle'_{A}$ unaffected.  The cancellation can \textit{only} happen for $\tilde{\Sigma}$ exchange, as follows from the \eqref{OPEAppendixGeneralOPEResult}.  Shadow operators have appeared previously in the dS/CFT literature; see \cite{Bzowski:2015pba,Anninos:2017eib}, for example.} to wavefunction coefficients and that this second OPE application is only a rough argument. In the end, after using \eqref{FourPointFunctionFromWU}, wavefunction heuristics give the same scaling result for the collapsed trispectrum as directly applying the OPE to \eqref{Collapsed4ptConfigurationResultInIn}.

\subsection{Soft Tensor Bispectrum}\label{app:softensbi}

We can apply similar methods to analyze the contribute of $\sigma$ to the following correlator:
\begin{align}
\lim _{\q\to 0}\langle \gamma_{\q}^\lambda\zeta^{\phantom{\lambda}}_{\k-\q/2}\zeta^{\phantom{\lambda}}_{-\k-\q/2}\rangle'\, .
\end{align}
However, in order for the answer to be non-trivial, we need to assume that the conformal symmetry is broken.  Applying the OPE \eqref{OPENeededForInIn} directly and keeping only the $\zeta\zeta\to \sigma$ channel contribution, one finds
 \begin{align}
\lim _{\q\to 0}\langle \gamma^\lambda_{\q}\zeta^{\phantom{\lambda}}_{\k-\q/2}\zeta^{\phantom{\lambda}}_{-\k-\q/2}\rangle'&\propto k^{t-4} \hat Y^\lambda_s(\theta,\varphi)\,\langle\gamma_\q^{\lambda}\sigma_{-\q}^{\lambda}\rangle'\,  ,\label{SoftGZZOPE}
 \end{align}
where we have expanded $\sigma_{i_1\cdots i_s}$ into the helicity basis and contracted the momenta, with the angles defined by $\cos\theta=\hat\k\cdot\hat \q$ and $\cos\varphi=\eps\cdot\hat\k$. The mixed two-point function is only non-trivial if the $\sigma$ field has a spin-2 component, corresponding to the restriction $t\le 1$. If the conformal symmetry is preserved, then $\langle\gamma_\q^{\lambda}\sigma_{-\q}^{\lambda}\rangle'=0$. However, if we assume that scale symmetry holds, but special conformal symmetries are softly broken, then it is possible to have a non-zero two-point function:
\begin{align}
	\langle\gamma_\q^{\lambda}\sigma_{-\q}^{\lambda}\rangle' = \frac{\rho}{q^{t+2}}\, ,\label{SCTBrokenCorrelator}
\end{align}
where $\rho$ is a small parameter characterizing the breaking of special conformal symmetry. Inserting \eqref{SCTBrokenCorrelator} into \eqref{SoftGZZOPE} and assuming that the OPE \eqref{OPENeededForInIn} holds up to $\mathcal{O}(\rho)$ corrections, we find the leading order scaling:
 \begin{align}
\lim _{\q\to 0} \langle\gamma_{\q}^\lambda\zeta^{\phantom{\lambda}}_{\k-\q/2}\zeta^{\phantom{\lambda}}_{-\k-\q/2}\rangle'\propto \frac{\rho}{q^3k^3}\left (\frac{q}{k}\right )^{1-t}\hat Y^\lambda_s(\theta,\varphi)\, , \quad t\le 1 \, ,\label{GZZFromOPE}
 \end{align}
consistent with \eqref{TSS}.  Similar results follow from a wavefunction analysis.

\newpage
\bibliographystyle{utphys}
\bibliography{HigherSpin}

\end{document}